\documentstyle[aps,prb,epsf,multicol]{revtex}

\title{Extreme Type-II Superconductors in a Magnetic Field: A Theory of
Critical Fluctuations}

\author{Zlatko Te{\v s}anovi{\' c}\cite{zbt}}
\address{Department of Physics and Astronomy, Johns Hopkins University,
Baltimore, MD 21218, USA
\\ {\rm(\today)}}
\begin{document} 
\maketitle
\begin{abstract}
A theory of critical fluctuations in extreme type-II superconductors
subjected to a finite but weak external magnetic field is presented.
It is shown that the standard Ginzburg-Landau representation of this
problem can be recast, with help of a novel mapping, as a 
theory of a new ``superconductor", in an effective magnetic 
field whose overall value is zero, consisting of the 
original uniform field and a set of
neutralizing unit fluxes attached to $N_{\Phi}$ fluctuating vortex lines.
The long distance behavior of this theory is governed by 
a phase transition line in the $(H,T)$ plane, $T_{\Phi}(H)$, along which
the new ``superconducting" order parameter, $\Phi ({\bf r})$,
attains long range order. 
Physically, this novel phase transition arises through
the proliferation,  or ``expansion", of thermally-generated 
infinite vortex loops in the background of field-induced vortex lines. 
Simultaneously, the field-induced vortex lines loose their
effective line tension relative to the field direction.
It is suggested that the critical behavior at $T_{\Phi}(H)$ belongs to
the universality class of the anisotropic
Higgs-Abelian gauge theory, with the original magnetic field
playing the role of ``charge" in this fictitious ``electrodynamics". 
At zero field, $\Phi ({\bf r})$ and the familiar superconducting order
parameter, $\Psi ({\bf r})$, are equivalent, and the effective line
tension of large loops and the helicity modulus vanish simultaneously,
at $T=T_{c0}$. In a finite field, however, these two forms of ``superconducting"
order are not the same and the ``superconducting" transition is 
generally split in two branches: the helicity modulus typically vanishes
at the vortex lattice melting line, $T_m(H)$, while the line tension and 
associated $\Phi$-order disappear only at $T_{\Phi}(H)$. 
We expect $T_{\Phi}(H)>T_m(H)$ at lower fields and
$T_{\Phi}(H)=T_m(H)$ for higher fields. Both $\Phi$- and $\Psi$-order are
present in the Abrikosov vortex lattice ($T<T_m(H)$) while both are absent
in the true normal state ($T>T_{\Phi}(H)$). 
The intermediate $\Phi$-ordered phase, between
$T_m(H)$ and $T_{\Phi}(H)$, contains precisely $N_{\Phi}$ field-induced
vortices having a finite line tension relative to ${\bf H}$
and could be viewed as a ``line liquid" in the long wavelength limit.
The consequences of this ``gauge theory" scenario for 
the critical behavior in high temperature
and other extreme type-II superconductors are explored in detail, with 
particular emphasis on the questions of 3D XY versus Landau level scaling,
physical nature of the vortex ``line liquid" and the true normal state,
and fluctuation thermodynamics and transport.
It is suggested that the empirically established ``decoupling transition"
may be associated with the loss of integrity of field-induced vortex
lines as their effective line tension disappears at $T_{\Phi}(H)$.
A ``minimal" set of requirements for 
the theory of vortex lattice melting
in the critical region is also proposed and discussed.
The mean-field based description of the melting
transition, containing 
only field-induced London vortices, is shown to be in violation of
such requirements.\\[0.2cm]
Pacs-numbers: 74.40.+k, 74.25.Bt, 74.20.De, 74.25.Dw, 74.25.Ha, 74.60.Ec
\end{abstract}
\noindent

\begin{multicols}{2}

\section{Introduction}
Recent intense activity in the area of superconducting fluctuations
has brought into sharp focus the 
following fundamental questions: What is the relationship between
the Landau level-based\cite{pierson,roulinlll,jeandupeux} 
and the 3D XY-based\cite{salamon,howson,moloni}
descriptions of superconducting fluctuations in magnetic field?
Can the mean-field based London
model containing only magnetic field-induced vortices\cite{dodgson}
describe the vortex lattice
melting transition in the region of strong (critical)
fluctuations? What is the nature of the normal
phase and can it be usefully represented 
as a ``line liquid"\cite{nelson} of field-induced
vortices? What role is played at finite fields by thermally-generated 
vortex loops,\cite{zt,sudboold} which are responsible 
for the zero-field transition in extreme type-II
superconductors? Particular importance and urgency has 
been attached to these questions
following the ground-breaking  experiments\cite{zeldov,schilling,roulin}
on the thermodynamics of vortex lattice melting transition\cite{gammel}
which clearly indicate that the low-field end of the melting line
is entering the critical regime of high temperature superconductors.

In this paper, precise answers to these questions are provided within 
a theoretical framework which allows for a systematic solution to the problem of
critical fluctuations in an extreme type-II superconductor subjected
to a finite, but weak magnetic field. This framework is built around
the ``gauge theory" scenario proposed earlier.\cite{zt}
Two main predictions follow from this scenario:\cite{zt} first, there is a new
transition line in the $H-T$ phase diagram, $T_{\Phi}(H)$, 
along which a thermally-generated vortex loop ``expansion"
takes place, reminiscent of the zero field transition. 
At $T_{\Phi}(H)$, well defined field-induced vortex lines are formed, having a {\em finite}
line tension {\em relative} to the field direction. Initially, these lines are in 
a liquid state and solidify only at some lower temperature $T_m(H)$ (Fig. 1). This is
different from the Abrikosov's theory, where vortices and 
their (Abrikosov) lattice are formed simultaneously, at $H_{c2}(T)$;
second, in contrast to the 3D XY behavior at zero field, the description of the
critical behavior along $T_{\Phi}(H)$ requires the combination of a 
complex ``superconducting" order parameter
$\Phi$ associated with vortex loops {\em and} a fictitious gauge field ${\bf S}$, 
describing fluctuations in the background system of field-induced vortex lines. 
The magnetic field determines the ``charge" which couples $\Phi$ and ${\bf S}$.
The physical picture arising from the ``gauge theory"
is remarkably detailed and compelling, and so entirely
distinct from the ``standard" approach \cite{blatterrmp},
that a concentrated effort should be directed at exploring its 
consequences.  The main purpose of this paper is to provide 
a novel explicit model for critical fluctuations, to examine
its main ramifications in some detail, and to advance
a set of specific predictions which can help establish the value of the
``gauge theory" description though experiments and numerical
simulations.

The essential feature of our description is that it contains, on equal footing,
{\em both} the field-induced {\em vortex lines} {\em and} thermally-generated
critical fluctuations of the superconducting order parameter $\Psi$, associated
primarily with {\em vortex loops}. It is the latter that 
dominate the entropy in the critical region.\cite{zt} 
This is in fundamental contrast to
other approaches which include {\em only} the field-induced vortices:
the Landau level (LL) description\cite{zta,yeo,ikeda,hu} (where other fluctuations become
irrelevant at sufficiently high fields) and the mean-field 
based picture of London vortices\cite{blatterrmp} 
(where other fluctuations can be ignored at sufficiently low temperatures,
far below critical regime).
Following the prescription proposed earlier \cite{zt}, which seeks to conveniently
isolate the background of field-induced from thermally-generated degrees of freedom, we
derive the following results: First, it is shown in Sec. II
that the familiar and frequently used ``helium" or ``London model"
of extreme type-II superconductors, in which the amplitude fluctuations
are suppressed, allows for a direct mapping of the original problem to that of
a new ``superconductor", whose order parameter $\Phi$ experiences
an overall magnetic field composed of the uniform external field, ${\bf H}$, and
the set of $N_{\Phi}$ neutralizing ``fluxes" attached to fluctuating vortex lines.
This novel mapping constitutes an explicit and transparent 
realization of the general connection proposed in Ref.\cite{zt}
The ``helium model" is then a candidate to,
in addition to the familiar vortex lattice melting line, $T_m(H)$, exhibit the
conjectured ``$\Phi$-transition" (or the vortex loop ``expansion" transition
in a {\em finite} field),
the universality class of which is 
defined by an anisotropic Higgs-Abelian gauge theory.\cite{zt,helium}
Physically, this $\Phi$-transition corresponds to vanishing of
the effective line tension for very large thermally-generated
vortex loops: at $T_{\Phi}(H)$, the energy-entropy balance in
the free energy shifts in favor of large loops and spontaneously created
infinite vortex-antivortex paths proliferate across the system.
The ensuing change in the topology of the vortex paths
results in the breaking of a ``dual" $U(1)$ symmetry and
a thermodynamic phase transition. Simultaneously, the field-induced 
vortex lines loose their line tension
relative to the field direction and the ``line liquid" description
breaks down.  As our second result, it is shown that the fictitious gauge theory
passes a crucial test, allowing us to 
connect its ``charge" to the original magnetic field.
Third, we use this connection in Sec. IV to construct scaling 
functions for the critical thermodynamics of extreme type-II
superconductors. Furthermore, the much-debated difference between the 3D XY-like
description at low fields and the 
LL description appropriate at high fields is closely linked here to
the difference between the extreme ``type-II" and the extreme ``type-I" behavior
of the gauge theory (Sec. III).
Fourth, it is shown in Sec. V that a vortex loop
``expansion" leads to an abrupt drop in the coefficient of the $q^2$-term in
the helicity modulus, from which one can extract the thermodynamic
exponent ($\nu$) of the $\Phi$-transition. Related criteria are also proposed
which test for the presence/absence of effective ``diffusion" of vortex lines
along the field and demonstrate close relation between the
``$\Phi$-order"\cite{zt} and 
viability of the vortex ``line liquid"\cite{nelson} description. 
These predictions, based only on global topological properties of
loops and lines, can be used to efficiently identify 
the vortex loop ``expansion" line, $T_{\Phi}(H)$,  
in numerical simulations of the weakly frustrated 3D XY and related models. 
Fifth, assuming widely used form of 
dynamical scaling, the explicit expression for the 
fluctuation conductivity, $\sigma (T,H)$, is
derived in Sec. VI in 
the vicinity of the $T_{\Phi}(H)$ line. At $T_{\Phi}(H)$ there
is an experimentally detectable 
rapid onset of additional dissipation, caused by thermally expanding vortex
loops whose size is reaching sample boundaries. It is tempting to
associate this onset at $T_{\Phi}(H)$ with what is empirically known as the
``decoupling" transition,\cite{decouplingi} although the physical origin of such
additional dissipation in our theory is entirely unrelated to any ``decoupling"
of any ``layers".\cite{decouplingii} Instead, it signifies the loss of integrity
of field-induced vortex lines as 
their effective line tension disappears at $T_{\Phi}(H)$.
At this point, one also expects a distinct change in the pinning properties
of the liquid state: there is no pinning in the true normal state
above $T_{\Phi}(H)$ just like there is
no pinning above $T_{c0}$. In this sense $T_{\Phi}(H)$ represents
an upper boundary for pinning and could be viewed loosely as
``renormalized" $H_{c2}(T)$.  Finally, in Sec. VII, it is 
demonstrated that the vortex lattice melting transition in the
critical region involves {\em simultaneous} ordering of the field-induced
{\em and} thermally-generated degrees of freedom and thus cannot be
faithfully represented by a
mean-field based London model,\cite{dodgson}
which includes only the former. Actually, as the melting line tends toward
$T_{c0}$ in the limit of vanishing magnetic field, 
the entropy change involved in ordering 
of thermally-generated loops overwhelms the 
configurational entropy of the field-induced vortex lines. This provides 
direct theoretical support for the fundamental nature and significance of the
experiments by Zeldov et al.,\cite{zeldov} Schilling et al.,\cite{schilling}
and Roulin et al.,\cite{roulin}
and new numerical simulations of Nguyen and Sudb{\o}\cite{sudbo,ryu}.

\section{From the Ginzburg-Landau theory to the Gauge theory}
The starting point is
the anisotropic Ginzburg-Landau (GL) theory
$Z=\int{\cal D}\Psi\exp\bigl \{-\int d^3r{\cal F}/T\bigr \}$,
where
\begin{equation}
{\cal F} =
\alpha\vert\Psi\vert ^2 + \sum_{\mu =\parallel,\perp}\gamma_{\mu}
\vert (\nabla_{\mu} + \frac{2ei}{c}{\bf A}_{\mu})\Psi\vert^2
+ \frac{\beta}{2}\vert\Psi\vert^4~~,
\label{ei}
\end{equation}
and $\alpha = a_0 (T-T_c)$, $\gamma_{\mu}$, and $\beta$ are
the GL coefficients.
Free (periodic) boundary conditions
are imposed in $\parallel$ ($\perp$) direction.
The limit $\kappa\to\infty$ is considered, which is particularly appropriate for 
high temperature superconductors (HTS).
In this limit, the external magnetic field, 
${\bf H}=\nabla\times{\bf A}_{\perp}$, acts as a {\em constraint}, forcing
every allowed configuration of the system to have the overall vorticity $N_{\Phi}$
along ${\bf H}$ ($\parallel {\bf\hat z}$). The overall vorticity along ${\bf H}$
is defined as a line integral $\int d{\bf l}\cdot\nabla\varphi /2\pi$,
where the contour of integration goes around perimeter of the system
in the xy-plane and $\varphi ({\bf r})$ is the phase of
$\Psi$.  $N_{\Phi}$, the number of
elementary flux quanta $\phi_0$, is given by $L_{\perp}^2/2\pi\ell ^2$, where
$\ell = \sqrt{c/2\vert e\vert H}$ is the magnetic length. 
It is assumed that this constraint 
is enforced by $N_{\Phi}$ vortex paths, 
meandering from one end of the system to
another, along ${\bf H}$.\cite{zt} 

	In this paper, a new method is introduced
to enforce the constraint {\em explicitly}, 
by considering a {\em different} partition function:
$Z'=\int{\cal D}\Phi\int\prod_{i=1}^{N_{\Phi}}
({\cal D}{\bf r}_i[s]/N_{\Phi}{\rm !})
\exp\bigl \{-\int d^3 r{\cal F}'/ T\bigl \}$, with
\begin{equation}
{\cal F}' = 
\alpha\vert\Phi\vert ^2 + \gamma_{\mu}
\vert (\nabla_{\mu} + i{\bf U}_{\mu} +\frac{2ei}{c}{\bf A}_{\mu})\Phi\vert^2
+ \frac{\beta}{2}\vert\Phi\vert^4.
\label{eii}
\end{equation}
$Z'$ describes the system of $N_{\Phi}$ ``shadow", or s, vortices 
$\{{\bf r}_i[s]\}$ in thermal equilibrium with
a complex field $\Phi ({\bf r})$. 
These s vortices sample arbitrary paths that 
originate (terminate) at $z=0$ ($z=L_{\parallel}$)
and differ from the ones introduced in Ref.\cite{zt} 
by the full inclusion of ``overhang" configurations. 
The effective magnetic field ${\bf H'}$ experienced by
$\Phi$ consists of the uniform external field ${\bf H}$ and the
collection of unit ``fluxes" attached to s vortices:
$\nabla\times {\bf U} = 2\pi {\bf n}_s({\bf r})$; 
$\nabla\cdot {\bf U} = 0$, 
where ${\bf n}_s({\bf r})$
is the flux density associated with a given configuration of s vortices,
$\{ {\bf r}_i[s_i]\}$:\cite{footo}
\begin{equation}
{\bf n}_s({\bf r}) = \sum_i^{N_{\Phi}}\int_{\cal L}d{\bf r}_i
\delta ({\bf r} - {\bf r}_i[s_i])~~~,
\label{es}
\end{equation}
with ${\cal L}$ denoting the line integral.
The net value of ${\bf H'}$ averaged
over the system {\em vanishes}.  

The superconductors (\ref{ei}) and (\ref{eii}) are {\em equivalent}
within the familiar ``helium model" of extreme type-II behavior; they
are just two different representations of the same physical problem.
To show this we recall the main features of the ``helium model":\cite{popov}
the true transition temperature $T_{c0}$ (Fig. 1) is assumed to be {\em sufficiently
below} the mean-field $T_c$ for amplitude fluctuations to have effectively
subsided. Around $T_{c0}$, the relevant fluctuations are considered to be those of 
London-type vortex loops/lines with steric repulsion and well-defined,
tight cores of size $a \ll \ell$. Consider now a single configuration
of these loops and lines. First, we extract the singular part of $\nabla\varphi ({\bf r})$ 
by solving two equations: $\nabla\times\nabla\varphi = 2\pi {\bf n}({\bf r})$ and
$\nabla\cdot\nabla\varphi =0$, where ${\bf n}({\bf r})$ is defined by the same
expression as ${\bf n}_s$ (\ref{es}) but with the summation running over
{\em all} vortex loops and lines. After this ``vortex" part 
has been extracted, the rest of $\Psi ({\bf r})$ is assumed to 
take the form which minimizes ${\cal F}$ for a given configuration 
of these line singularities. We then integrate over all regular (``spin-wave")
fluctuations in $\varphi ({\bf r})$.  Finally, all such distinct 
configurations of vortex loops and lines are 
summed over to produce $Z$ (\ref{ei}).
This same procedure is imposed on $\Phi ({\bf r})$: first we extract the part 
of {\em its} phase, $\nabla\phi ({\bf r})$,
due to vortex loop/line singularities in $\Phi ({\bf r})$ and then
determine the rest of $\Phi ({\bf r})$ by minimizing ${\cal F}'$ (\ref{eii})
for a given configuration of these line defects {\em and} 
s vortices.  Again, we integrate over all ``spin-wave" fluctuations of $\phi$
relative to this given configuration of loops and lines.
By direct comparison of these ``helium model" expressions
obtained from (\ref{ei}) and (\ref{eii}), 
it is evident that all configurations contributing 
to the original $Z$ are reproduced in $Z'$ 
and have the same energy.
However, some of these configurations are counted more than once in $Z'$.
This overcounting of configurations in $Z'$ relative
to $Z$, given by $(N_{\Phi} + N_a){\rm !}/N_{\Phi}{\rm !}N_a{\rm !}$ with
$N_a$ being the number
of vortex lines in $\Phi$ which traverse the sample along ${\bf H}$,
is a surface effect in 3D and should be unimportant in the thermodynamic limit,
$L_{\perp}$, $L_{\parallel}\to \infty$.
Moreover, within the conjectured $\Phi$-ordered phase (Fig. 1),
the configurations with $N_a \not =0$ are irrelevant in the
thermodynamic limit and there is no overcounting at all.
We conclude that, within the ``helium" model, the free energy evaluated from $Z'$
coincides with the free energy of the original problem (\ref{ei}) and the two 
superconductors have {\em identical} thermodynamics.\cite{footi} 
Consequently, Eq. (\ref{eii}) accomplishes a 
straightforward and transparent reformulation of the original problem,
in the spirit of Ref.\cite{zt},
while avoiding more cumbersome gauge transformation method
employed there.\cite{footo} More details on the ``helium model" are
presented in Appendix A.

If we relax the above minimization condition on the amplitude of our order parameters,
$\Psi$ and $\Phi$, and permit weak amplitude fluctuations, we expect that
the above close relation between $Z$ (\ref{ei}) and $Z'$ (\ref{eii}) still holds,
as long as the parameters of the GL theory keep us in the extreme type-II limit.
This requires the {\em average} core size $a$ to be smaller than the 
average spacing between vortex segments, so that vortex excitations
remain well-defined. It is precisely this same requirement
that is invoked to justify the frequent use of the ``helium model" to emulate fluctuation
behavior of extreme type-II superconductors in zero field. It is natural
to expect that, if such requirement is satisfied at zero field, it will remain
so at low fields, such that $a\ll\ell$.
Based on this, on the equivalence of representations (\ref{ei}) and (\ref{eii})
in the ``helium model" limit, and on our general expectation that the extreme
type-II behavior with only weak amplitude fluctuations is effectively equivalent to
the ``helium model", for the rest of this 
paper I consider (\ref{eii}) to be simply an alternative 
formulation of the original problem. This new reformulation (\ref{eii})
can now be used instead of (\ref{ei}) to compute various fluctuation
properties and, most importantly, its
{\em critical} behavior should {\em coincide} 
with that of the original GL theory (\ref{ei}).

The advantage of $Z'$ (\ref{eii}) is that, by 
isolating the background of field-induced degrees of freedom (s vortices),
it focuses our attention on the new ``superconducting"
order parameter, $\Phi ({\bf r})$, and its spatial correlations, measured
by $\langle\Phi ({\bf r})\Phi ^*({\bf r'})\rangle$, where $\langle\cdots\rangle$
denotes thermal average over $Z'$. All excitations of $\Phi ({\bf r})$
are {\em thermally generated}, in the following precise sense: {\em every configuration
of $\Phi ({\bf r})$, contributing a finite weight to $Z'$ in the thermodynamic limit,
has the overall vorticity along ${\bf H}$ equal to zero}. In particular,
$\Phi ({\bf r})$ contains vortex loop excitations, whose ``expansion" across the system
is the mechanism behind the $H=0$ superconducting transition (Fig. 2). 
By focusing on $\Phi$, we can fashion a theory of the
strongly {\em interacting} Wilson-Fisher (3D XY)  critical point, ``perturbed"
by a weak field.\cite{zt} This is precisely the opposite of the classic approach,\cite{bnt}
where the Gaussian theory in a {\em finite} field is perturbed by
weak interaction. Such approach starts with the LL structure from the
outset and its critical behavior is always dominated by the lowest LL.\cite{bnt,zta}
In the new formulation (\ref{eii}) we had built in from the start our expectation that
the weak field modifies zero field configurations only by introducing 
a low density of s vortex lines, the cores of which are well-defined by virtue of
strong amplitude correlations at the 3D XY critical point. This is a ``low-field"
approach by design and offers a better prospect of constructing the desired theory.

To extract such a theory from (\ref{eii}) we must resort to
approximations. We construct the long wavelength ($\gg\ell$) limit
of (\ref{eii}) by coarse-graining vorticity fluctuations produced
by s vortices. The ``hydrodynamic" vorticity ${\bf V}({\bf r})$ is 
defined as the coarse-grained version
of the ``microscopic" flux density $\Delta {\bf n}_s({\bf r})=
{\bf n}_s({\bf r})-(2\pi\ell ^2)^{-1}{\bf z}$. 
Upon inserting $\int {\cal D}{\bf V}\delta [{\bf V}({\bf r})
-\Delta {\bf n}_s({\bf r})]$ in (\ref{eii}), integrating over
$\{ {\bf r}_i[s]\}$, and after 
introducing the fictitious vector potential 
$\nabla\times {\bf S} = 2\pi {\bf V}$; $\nabla\cdot {\bf S}=0$,
the effective long wavelength theory becomes
$\sim\int {\cal D}\Phi\int {\cal D}{\bf S}
\exp \bigl \{-\int d^3 r{\cal F}_{\rm eff}/T\bigr \}$,
\begin{equation}
{\cal F}_{\rm eff} = 
\alpha\vert\Phi\vert ^2 + \gamma_{\mu}
\vert D_{\mu}\Phi\vert^2
+ \frac{\beta}{2}\vert\Phi\vert^4 + \frac{K_{\mu}}{2}(\nabla\times{\bf S})_{\mu}^2~,
\label{eiii}
\end{equation}
where $D_{\mu} =\nabla_{\mu} + i{\bf S}_{\mu}$ and
$K_{\perp} (T,H)
=c_{\perp}\Gamma ^{-1}T\ell$,
$K_{\parallel}(T,H)
=c_{\parallel}\Gamma T\ell$. Higher powers and derivatives of
$(\nabla\times {\bf S})_{\mu}$, essential for the description of the
vortex lattice melting,  also appear in (\ref{eiii}), but are unimportant
at $T_{\Phi}(H)$.
Detailed derivation of the ``gauge theory"
(\ref{eiii}) is given in Appendix A.
$c_{\perp,\parallel}\sim {\cal O}(1)$ are dimensionless
and have a relatively weak $H,T$ dependence in that portion of the critical region
which is well described by Eq. (\ref{eiii}).\cite{zt} 
A close relation between $K_{\perp,\parallel}$, $c_{\perp,\parallel}$
and the components of the 
helicity modulus tensor of the GL theory (\ref{ei}) is discussed later in the
text (see Sec. V and the Appendix B).  $\Gamma$ is the anisotropy at $T_{c0}$.
Small $H$-dependent corrections to GL coefficients that also should appear
in (\ref{eiii}) are ignored, since they are not important for our present purposes.

The following assumptions have been used in going 
from (\ref{eii}) to (\ref{eiii}) (see also Appendix A):\par
\noindent
i) The correlation length, $\xi_{\Phi}$, associated
with the new order parameter $\Phi$, is not limited by $\ell$ and can be
much longer than the original 
superconducting correlation length, $\xi_{\rm sc}$, associated with $\Psi$. 
Of course, this is the basic reason why we are interested in the reformulation
(\ref{eii}) in the first place. When $\xi_{\Phi}\gg\ell,\xi_{\rm sc}$, 
this assumption enables us to drop as irrelevant\cite{zt} at long distances,
terms containing higher derivatives and powers of ${\bf S}$ from Eq. (\ref{eiii}). 
Note, however,  that such higher order terms in (\ref{eiii}), 
particularly those reflecting the absence of up-down symmetry along 
${\bf H}$ ($(\nabla\times {\bf S})_{\parallel}^3$ and the
like), must be restored when discussing vortex lattice melting (Sec. VII).
The gauge theory (\ref{eiii}) offers in this case ($\xi_{\Phi}\gg\ell,\xi_{\rm sc}$)
a direct access to the deeply {\em non-perturbative} 
regime of the original GL theory (\ref{ei}),
characterized by $\xi/\ell \gg 1$, where $\xi$ is the $H=0$
correlation length. In the opposite case, $\xi_{\Phi} \ll\ell$,
we are in the {\em perturbative} regime, $\xi/\ell \ll 1$, of the original theory.
The long wavelength expansion that led from (\ref{eii}) to (\ref{eiii})
is then not justified and the new reformulation (\ref{eii}) is not particularly useful. 
\par
\noindent
ii) The system (\ref{ei}) is not in its superconducting state
in the vicinity of the putative $\Phi$-transition.
This assumption fixes the form of the last two terms in (\ref{eiii})
(see Sec. V and the Appendix B).
Physically, it means that the system of s vortices contains configurations
that wind from one end of a sample to another in the $\perp$ directions
(xy-plane). The presence of such windings allows complete ``screening" of
an arbitrary infinitesimal field ${\bf h}({\bf r})$ added to ${\bf H}$ and
the helicity modulus tensor
vanishes along {\em all} of its principal axes (see Sec. V and the Appendix B). 
This assumption has a strong theoretical justification.\cite{nordborg}
It must be emphasized, however, that, to my knowledge, there
is no rigorous argument which could rule out another possibility,
that of the state right below the $\Phi$-transition being an extremely anisotropic
``superconducting" liquid, containing no windings in the xy plane, with
a finite helicity modulus {\em along} the field and zero perpendicular to it. 
Indeed, some numerical studies are suggestive of this second possibility.\cite{teitel,ryu}
However, other numerical simulations\cite{sudbo,koshelev,tachiki}, 
as well as the available experimental data, favor our original
non-superconducting liquid assumption. 
{\em Both} alternatives can be described within the framework of the gauge theory,
with $K_{\perp}/K_{\parallel}\to\infty$ and potentially finite 
``mass terms" 
below $T_{\phi}(H)$ added to (\ref{eiii}) in the extreme anisotropy case. 
On general physical grounds,\cite{nordborg} 
I have chosen to explore in this paper the case
of finite anisotropy ratio $K_{\perp}/K_{\parallel}$ and vanishing mass terms
but the reader should be aware that the extreme anisotropy alternative 
remains a possibility\cite{footip} and would lead to results which, while similar
on general level, differ in details.\cite{footipp}
\par
\noindent
iii) There are two relevant lengthscales controlling
the critical behavior: $\xi_{\Phi}$, which characterizes {\em both} the
spatial correlations of $\Phi$ {\em and} the size of ``overhangs" in the
system of s vortices, and $\ell$, which characterizes the long wavelength fluctuations of
the background field-induced vorticity.
\par
\noindent
iv) The core effects can be ignored. Clearly, the ``helium
model" itself is perfectly well defined in the limit $a\to 0$. For
$a$ small but finite, there is a small correction to the core
line energy $\varepsilon_c\to\varepsilon_c + w_c{\bf H}\cdot{\bf v}$,
where ${\bf v}=d{\bf r}/ds$ is the ``velocity" of a vortex segment
and $w_c H/\varepsilon_c\sim a^2/\ell ^2 \ll 1$. Similarly,
there are ``velocity" dependent corrections to the short range
repulsion between vortex cores. Such terms are irrelevant
since they result in higher order derivatives in (\ref{eiii}). For example,
it is easy to see that the correction to the core energy cancels out
for any finite vortex loop and can be factored out for s vortices.

What is the physics behind gauge theory (\ref{eiii})?
The external field has been eliminated from the gradient terms in (\ref{eiii})
($\langle {\bf H'}\rangle =0$) but, of course, it has not vanished: 
it {\em reappears} through the $H$-dependence of $K_{\perp,\parallel}$. 
The gauge theory (\ref{eiii}) can be viewed as fictitious, anisotropic
``electrodynamics" with ``magnetic permeability" $\mu_0 = 1/4\pi T$.
The ``vector potential" ${\bf S}$ is coupled to
the``matter" field $\Phi$ via ``electrical charge" 
$${\tilde e}^2_{\perp,\parallel}
=\frac{\Gamma ^{1/3}}{c_{\perp,\parallel}\ell}\propto \sqrt{H}~~. $$
The above ``charge" and $K_{\perp,\parallel}$ describe
the ``polarizability" of the medium composed of s vortices and are directly related to
the long wavelength components of the helicity modulus tensor (Sec. V and the Appendix B).
This picture embodies the physical idea that the dominant effect of weak magnetic field
in (\ref{ei}), once $\xi_{\rm sc}$ has saturated to $\sim\ell$,
arises through ``screening" of large thermally-generated
loops by the background of field-induced vorticity, at distances $\gg\ell$.
Such ``screening" reduces the effective line tension of these large loops relative
to its value at the $H=0$ ($\tilde e_{\perp,\parallel} =0$) transition.
The strength of the ``screening" is measured by the fictitious ``Ginzburg parameter"
of (\ref{eiii}), 
\begin{equation}
\kappa_s^2 \sim c \frac{\beta \ell}{2a_0^2T_c\xi_{GL}^4}= 
\frac{b}{2q^2_0}\propto \frac{1}{\sqrt{H}}~~~,
\label{eiv}
\end{equation}
where $c  = (c_{\perp}^2c_{\parallel})^{1/3}$ and
$\xi_{GL}  = (\xi_{GL\perp}^2\xi_{GL\parallel})^{1/3}$, with
$\xi_{GL\perp,\parallel}=\sqrt{\gamma_{\perp,\parallel}/a_0T_c}$
being the GL coherence lengths. 
$b = \beta/a_0^2T_c\xi_{GL}^3$ and
$q_0^2 = {\tilde e}^2\xi_{GL}$ are the dimensionless
quartic coupling and ``charge", respectively.  
As $H\to 0$, the fictitious ``charge" vanishes and 
we recover the zero-field 3D XY critical point. For $H$ finite but weak,
the ``screening" is weak ($\kappa_s\gg 1$), indicating 
that the effects of finite ${\tilde e}$ are small compared to strong
amplitude correlations produced by the quartic term in the GL theory (\ref{ei}).
We have therefore manufactured a critical theory (\ref{eiii}) describing
the strongly-interacting Wilson-Fisher (3D XY) critical point weakly ``perturbed" by a finite
magnetic field (finite ``charge" $\tilde e_{\perp,\parallel}$).\cite{footiii}

The gauge theory scenario is clearly different 
from what takes place in spin systems, where the external field
couples {\em  paramagnetically} to the order parameter.
In an extreme type-II  superconductor (\ref{ei}), the {\em diamagnetic} coupling
of ${\bf H}$ to $\Psi$ does not explicitly break the $U(1)$ symmetry, which was spontaneously
broken at the zero-field 3D XY critical point. The high temperature phase
(true normal state) still retains the full $U(1)$ symmetry. This symmetry can be broken
at low temperatures,
either in a ``simple" way, with $\Psi$ acting as the order parameter, as is
the case in the ``vortex solid" state, or in a more
subtle fashion, with $\Phi$ assuming the role of the new order parameter.
Similarly, the gauge theory (\ref{eiii}) differs from frequently
used ``dimensional reduction" approaches,\cite{connor}
where the behavior of (\ref{ei}) at finite
fields is related to that at {\em zero} field but in a
{\em finite} system, the size of which is set by the magnetic length, $\ell$.
A typical dimensional reduction ($D\to D-2$) approach leads to the superconducting
correlation length which is limited by $\ell$, i.e. the ``system size".
This {\em agrees} with the gauge theory scenario,\cite{zt} since ``electrodynamics"
(\ref{eiii}) also predicts $\xi_{\rm sc} (H)\sim\ell$ in the critical
region (see Sec. V).  However, a dimensional reduction approach 
also predicts that {\em all other}
correlations are limited by $\ell$, and, consequently,
eliminates the possibility of any true thermodynamic phase transition
in the GL theory (\ref{ei}). This is in contradiction with the overwhelming
experimental and numerical evidence indicating some form of a ``vortex liquid"
to ``vortex solid" transition at low temperatures. In sharp contrast, 
gauge theory (\ref{eiii}) and reformulation (\ref{eii}) are fully
three-dimensional theories, just like (\ref{ei}). They naturally lead to
two basic types of correlations that can extend over distances $\gg\ell$
and produce phase transition(s) at low temperatures (Fig. 1):
those associated with positional order of s vortices and the familiar
superconducting order parameter $\Psi ({\bf r})$ and those associated
with the new ``superconducting" order parameter $\Phi ({\bf r})$.

The conjecture\cite{zt} that connects the critical behavior of 
an extreme type-II superconductor (\ref{ei}) to a fictitious 
superconductor in {\em zero} field (\ref{eiii}),
the ``charge" of which is set by the original external field $H$, must pass the following
test: the way $H$ enters in ${\cal F}_{\rm eff}$ must be consistent
with its being a relevant operator of scaling dimension 2 in the renormalization
group (RG) sense at the 3D XY critical point. This scaling 
dimension is suggested by dimensional analysis\cite{ffh} 
(relevant effects of $H$ enter through the dimensionless
ratio $\xi_{\rm sc}^2(H=0)/\ell ^2\propto 
H\xi_{\rm sc}^2(H=0)$), is correct to two-loop order\cite{lawrieii}
and is likely an exact property of the original GL theory (\ref{ei})
by virtue of gauge invariance. In addition, the scaling 
dimension appears independent on the nature of the zero-field critical
point (i.e., whether it is 3D XY or Gaussian). On the other hand,
as the ``charge" ${\tilde e}$ is turned on
in the gauge theory (\ref{eiii}), the RG analysis indicates that first,
the finite charge anisotropy (${\tilde e}_{\perp}\not = {\tilde e}_{\parallel}$) is
{\em marginally irrelevant},\cite{lubensky,herbut} 
and second, the scaling dimension of ``charge"
at the 3D XY critical point is 1/2, i.e., the relevant dimensionless operator
is ${\tilde e}\sqrt{\xi_{\rm sc}({\tilde e}=0)}$.
The second statement is {\em exact} to all orders in
perturbative RG and is also independent on the nature of the {\em neutral}
critical point.\cite{herbutzt} Since, in ${\cal F}_{\rm eff}$ (\ref{eiii}),
${\tilde e}^2\propto 1/\ell\propto\sqrt{H}$ this translates immediately to the
scaling dimension of $H$ being 2, as required. More generally,
for dimension $D<4$, ${\tilde e}^2\propto \ell ^{D-4}$, while the scaling
dimension of ${\tilde e}$ is $2 - (D/2)$, again consistent with
the scaling dimension of $H$ being 2.  Note that in both formulations,
Eqs. (\ref{ei}) and (\ref{eiii}), the corresponding relevant operators $H$ and 
${\tilde e}$ ($\propto H^{1/4}$) are protected against acquiring anomalous dimensions by the
same symmetry, the gauge invariance.
These results demonstrate internal consistency of the 
coarse-graining procedure leading to (\ref{eiii})
and strongly support the conjecture\cite{zt} that ${\cal F}_{{\rm eff}}$
captures the long wavelength (critical) behavior of (\ref{eii}) and (\ref{ei}).
In what follows, I promote this conjecture to a fact and examine its consequences.

\section{High fields versus low fields}
Two key consequences for the physics of the present problem follow 
from ${\cal F}_{\rm eff}$. First, the gauge theory (\ref{eiii})
has two distinct regimes of behavior: the weak ``screening" limit ($\kappa_s\gg 1$)
corresponding to the extreme ``type-II" limit of the fictitious ``electrodynamics"
and the strong ``screening" limit ($\kappa_s\ll 1$) corresponding to the
extreme ``type-I" behavior. The extreme ``type-II" behavior of (\ref{eiii}) 
is precisely the low-field
regime of the original theory (\ref{ei}) which exhibits the 3D XY-like 
critical fluctuations. In this low-field regime, the ``screening" provided
by the background of field-induced vorticity is weak and the dominant
fluctuations are still London-type vortex loops and lines. The core
size $a$ remains small and well-defined, 
kept in check by strong amplitude correlations coming from the 
quartic term in (\ref{ei}), just
as was the case at the zero-field 3D XY critical point. 
It is in this sense
($\kappa_s\gg 1$) that we can think of a 3D XY critical 
point weakly ``perturbed" by a finite field.\cite{footiii} 
In the extreme ``type-I" limit, the situation is entirely different. There,
the ``screening" is strong ($\kappa_s\ll 1$) and the amplitude fluctuations
ran rampant. It is not possible any longer to think of relevant fluctuations
in the gauge theory (\ref{eiii}), nor in (\ref{eii}) and (\ref{ei}), as being
London-like vortices. Rather, amplitude fluctuations are now of essential 
importance and individual vortex cores are ill-defined. In the gauge
theory (\ref{eiii}), the two regimes are separated by the condition
$\kappa_s \sim 1$. However, a word of caution must be inserted here since,
once we are in the ``type-I" regime of (\ref{eiii}),
our original line of reasoning that lead from Eq. (\ref{ei}) to
the gauge theory (\ref{eiii}) via reformulation (\ref{eii}), is itself compromised
and it is not clear whether there is a useful connection between the extreme
``type-I" limit of (\ref{eiii}) and our original problem (\ref{ei}).
Instead, we must return back to the beginning (\ref{ei}) and start
from scratch. It is natural to identify this extreme ``type-I" behavior at high fields,
characterized by strong amplitude fluctuations, as the regime in which
the Landau level structure of the original GL theory (\ref{ei}) becomes important.
The condition $\kappa_s\sim 0.4/\sqrt{2}$,\cite{herbutzt} 
separating ``type-II" from ``type-I"
``electrodynamics" in Eq. (\ref{eiii}), translates to the criterion for the
external magnetic field, $H\sim H_s$, telling us whether $H$ is ``low" or ``high".
From Eq. (\ref{eiv}) one gets:
\begin{equation}
H_s \cong\bigl(\frac{c}{0.16}\bigr)^2
b^2H_{c2}^{GL}(0)~~~.
\label{ev}
\end{equation}
If $H\ll H_s$ then the field is ``low" and the use of 
a 3D XY-like description is justified.
In the opposite limit, $H\gg H_s$, the field is ``high" and a 3D XY-like 
description falls apart (Fig. 1). If this is the case, we must abandon our zero-
and low-field imagery of the 
``helium model" and use as a starting point an approach that is explicitly
designed to deal with a high-field behavior, an example being the
GL-LLL theory.\cite{zta,hu} Note that $H_s$, within factors of order unity,
{\em coincides} with $H_b$, the field below
which the high-field, Landau level-based description breaks down, due
to strong LL mixing.\cite{zta}. Since this criterion\cite{zta} is derived from
entirely different arguments, we briefly reproduce it here for completeness.
Going back to the GL theory (\ref{ei}), we expand $\Psi ({\bf r}) =
\sum_{j=0}\Psi _j({\bf r})$ in the set of LL manifolds, $\Psi _j({\bf r})$.
Recast in terms of dimensionless variables,
the GL free energy functional becomes:
\begin{equation}
\int d^3 r\bigl \{\sum_{j=0}\bigl[t+(2j+1)h\bigr]\vert\Psi _j\vert^2 + 
\vert\nabla_{\parallel}\Psi _j\vert^2 + \frac{b}{2}\vert\Psi\vert^4\bigr \},
\label{evi}
\end{equation}
where $t=(T/T_c)-1$, $h=H/H_{c2}^{GL}(0)$
and $b$ is defined below Eq. (\ref{eiv}). After rescaling $\Psi$ and
${\bf r}$ by $b$ in such a way that the coefficients of 
quartic and gradient terms in (\ref{evi}) are set to 1/2 and 1,
respectively, the ``mass term" for $\Psi _j$ becomes
$$\frac{t}{b^2}+(2j+1)\frac{h}{b^2}~~~.$$
As we reduce the field $H$, for $T$ in the critical region ($T\approx T_c$), 
the mixing of LLs becomes strong when $h/b^2$ becomes some number of
order unity. We can view this as a ``Ginzburg criterion" along 
the $H$-axis. It implies that the high-field, Landau level description 
becomes inadequate for fields less than:
\begin{equation}
H_b \sim b^2 H_{c2}^{GL}(0)\sim
{\rm Gi}H_{c2}^{GL}(0)~~~,
\label{evii}
\end{equation}
where we have used a close relation
between $b$ and a conventional Ginzburg fluctuation parameter, Gi.\cite{zta}
While the definition and meaning of Gi exhibit wide variations 
in the literature, typically Gi $\sim b^2$.\cite{blatterrmp}
As advertised, $H_b\sim H_s$. The same situation is encountered in 2D,
except now $H_s\sim H_b \sim b H_{c2}^{GL}(0)$.
The fact that the criterion for the breakdown
of the Landau level-based theory derived from the high-field side 
agrees with the region of validity of our 3D XY-like approach
derived from the opposite, low-field side,
is another argument in favor of the gauge theory scenario.

	While it is the GL theory (\ref{ei}) that provides a realistic description
of fluctuation behavior in extreme type-II superconductors, many numerical
studies are performed on the 3D XY model. The ultimate low-field critical behavior should
be the same and computational effort is much reduced. 
It is therefore useful to discuss here the physical meaning of
the ``high" and ``low" field regimes in the context
of the frustrated 3D XY model. There is an immediate difference between this model
and the GL theory (\ref{ei}) regarding the high field behavior. In the GL theory
this regime is dominated by Landau levels and is characterized by
strong amplitude fluctuations. In contrast, in the 3D XY model, the amplitude fluctuations
are frozen at the ``microscopic" level of a single XY spin. As a result,
there is no Landau level formation in this model. Instead,
the high field behavior of a uniformly frustrated 3D XY model, as one approaches
the ``mean-field" $H_{c2}(0)$, is entirely determined by the pinning
of field-induced vortices by the underlying lattice.  We can think of this
situation, to some extent, as having the LL structure of Eq. (\ref{evi})
thoroughly ``mixed" by a very strong external periodic potential.
There is, however, a relationship between the low-field critical behavior of the
3D XY model and the GL theory. It derives from our concept of ``screening" of
large thermally-generated vortex loops by the background of field-induced
vorticity. In the weakly frustrated 3D XY model such ``screening" is measured
by a parameter $\kappa _{sXY}$, which is the XY model counterpart of 
$\kappa_s$ in the GL theory (\ref{eiv}):
\begin{equation}
\kappa_{sXY}^2\sim \sqrt{\frac{f^{T}(T,H)}{f_{\Phi}}}\propto
\frac{1}{\sqrt{H}}~~~,
\label{exy}
\end{equation}
where $f_{\Phi}$ measures the uniform frustration and 
is the fraction of the elementary flux quantum $\phi_0$ per plaquette, while
$f^{T}(T,H)$ is the average density per plaquette
of vortex and antivortex segments $\parallel$ ${\bf H}$ piercing the xy-plane.
Note that $f^T(T,H)$ includes {\em all} such vortex segments, not just those
connected to infinite vortex loops. In order for the system to be in the
low-field critical regime of a weakly frustrated 3D XY model we need
$\kappa_{sXY}\gg 1$ or $f_{\Phi}\ll f^T(T,H)$. 

	The actual value of $H_s$ (or $H_b$) in high temperature superconductors is
of considerable importance. There are numerous estimates in the
literature, based both on Eq. (\ref{evii}) and on the analysis
of various experimentally measured quantities in terms of
either the GL-LLL theory or the so-called ``3D XY scaling" (see 
Sec. IV).
A direct estimate from Eq. (\ref{evii}), 
in a moderately anisotropic HTS system like optimally-doped YBCO, 
uses Gi $\approx 0.01$ and
$H_{c2}^{GL}(0)\approx 160$ Tesla, leading to $H_s\sim 1-2$ Tesla.
This estimate is subject to an irksome uncertainty, both intrinsic
(due to our inability to theoretically determine $H_s$ or $H_b$ with a precision
better than within factors of order unity) and extrinsic (due to
difficulties in extracting precise values of the GL parameters 
entering Eq. (\ref{ei}), although the situation here is rapidly
improving\cite{pierson}). The estimates of $H_b$ based on the fits of fluctuation
thermodynamics to the GL-LLL theory are in general agreement with
the above value of 1-2 Tesla\cite{pierson} and seem to give an upper limit
$H_b < 8$ Tesla.\cite{lawriei} Similar analyses, based on the fits to a low-field
``3D XY scaling", generally produce results which seem consistent
with the 3D XY-like behavior to much higher fields, 
14 Tesla or even higher.\cite{howson,moloni,friesen} An important difference 
between the two approaches is that, within the
GL-LLL theory, not only the scaling law but the scaling function
and explicit expressions for thermodynamic quantities 
are known with considerable accuracy.\cite{zta,pierson,hu}
In the 3D XY approach only the scaling law itself is known but the 
actual scaling function and, more importantly, the physics behind it are not.
The gauge theory scenario should help remedy this situation.

\section{Critical thermodynamics and $\Phi$-transition}
This brings us 
to the second important consequence of description (\ref{eiii}),
which has bearing on
the nature of critical behavior in the low-field ($H \ll H_s$), 
extreme ``type-II" limit of the gauge theory.
The most significant property in this regime   
is that, for ${\tilde e}$ small but {\em finite}, there is a true thermodynamic
phase transition separating the high and low temperature phases of the theory,
the ``normal" and the ``Meissner" state, respectively.
For ${\tilde e}=0$ this is the standard $H=0$ phase transition of the
Ginzburg-Landau theory. This phase transition is continuous and in the
universality class of the 3D XY model. The actual mechanism of the phase
transition is directly tied to the expansion of thermally-generated vortex
loops, as depicted in Fig. 2. 
In the ordered state below $T_{c0}$, there is a {\em finite} average
size for such loops, $\Lambda_{\Phi}$, 
and configurations which contain infinite loops, ``percolating"
from one end of the system to another, do not contribute to the partition
function in the thermodynamic limit. At distances much larger than $\Lambda_{\Phi}$,
there is nothing to disturb the long range correlations in
$\langle\Phi (0)\Phi ^*({\bf r})\rangle$: it is not possible to ``polarize"
closed loops at such large distances and they behave as bound ``dipoles". This
is what enables the long range phase order that characterizes
the superconducting state.  Above $T_{c0}$, as more and more
vortex segments are created by thermal excitation, the loops connect,
in the sense that now there is finite contribution to the partition 
function from configurations having infinite loops, ``percolating"
across the system. This implies that $\Lambda_{\Phi}\to\infty$ and
it is now possible to ``polarize" the system of loops over arbitrary
large distances. Such infinite vortex loops act as ``free charges"
and produce ``metallic screening" of small external magnetic fields,
resulting in vanishing of the helicity modulus, as discussed
in the next section. This picture of the 3D XY phase
transition as vortex loop ``expansion" has its origins in the works
by Onsager\cite{onsager,sudbo} and Feynman\cite{feynman},
in the context of superfluid helium,\cite{helium} but should equally well apply to
high temperature superconductors (HTS) with their short BCS coherence
lengths and extremely large $\kappa$ ($\sim 100$).\cite{schneider} 

As finite $\tilde e$ (finite $H$ in (\ref{ei})) is turned on in
Eq. (\ref{eiii}) we are facing potentially dramatic change in this picture.
In the neutral-superfluid picture described previously, vortex loops have
long range London-Biot-Savart interactions. Once $\tilde e$ is finite,
these interactions are ``screened" by the vector potential ${\bf S}$
and, at distances much longer than the ``penetration depth"
$\lambda_s \propto 1/{\tilde e}$,
all the interactions are short-ranged. The simplest and best known example
of this is just an ordinary superconductor at {\em zero} external field.
There $\tilde e$ is the real electrical charge $e$, while
${\bf S}$ turns into the ordinary Maxwell's vector potential ${\bf A}$.
This charged-superfluid problem has been studied extensively,
starting with Ref.\cite{halperin}, and is presently thought to have the following
properties\cite{herbutzt}: as already indicated in Sec. II,
the charge $e$ is a relevant operator in the RG sense, with scaling dimension
equal to $1/2$. This immediately destabilizes the neutral-superfluid, 
zero-charge 3D XY
critical point. There are, however, two new critical points,
characterized by {\em finite} charge. The behavior of strongly type-II
superconductors ($\kappa\gg 1$) is determined
by the stable critical point and describes the
{\em continuous} phase transition between the normal state and the Meissner
phase in real superconductors.\cite{herbutzt} Another critical point is {\em tricritical}
and unstable in one RG direction, in addition to temperature. This
{\em tricritical} point defines the transition between type-II (small
charge) and type-I (large charge) behavior and takes place for
$\kappa\sim 0.4/\sqrt{2}$.\cite{herbutzt} In a type-I superconductor, the
phase transition is expected to be {\em discontinuous}, as originally
argued in Ref.\cite{halperin}. In the type-II regime, where the transition
is continuous, the universality class for the charged-superfluid appears
to be the ``inverted 3D XY"\cite{dasgupta,kleinert}, or very ``close"
to it\cite{herbutzt,olsson} (see Appendix A for further details). 

What is the connection between these general properties of charged-superfluid model
and our problem? It stems from  the gauge theory (\ref{eiii}). This
theory looks just like the theory for charged-superfluid, except for the charge
anisotropy, which should be irrelevant\cite{lubensky,herbut}. 
The underlying physics, of course, is very different. There is no fluctuating
electrodynamic vector  potential in our case, since we are in the
$\kappa\to\infty$ limit. Instead, our fictitious vector potential
${\bf S}$ describes the long wavelength vorticity fluctuations in the background
s vortex system and our ``charge" $\tilde e$ is the original magnetic field
$H$ in disguise ($\tilde e^2\propto \sqrt{H}$). 
Despite this difference in physical meaning, the long distance
behavior of (\ref{eiii}) should still be closely related to the electrodynamics
of charged-superfluid. In particular, we expect two different thermodynamic
phases of (\ref{eiii}): the high temperature phase with 
only short range correlations in $\langle\Phi (0)\Phi ^*({\bf r})\rangle$
($\langle\Phi\rangle = 0$) and the low temperature phase, in which
$\langle\Phi (0)\Phi ^*({\bf r})\rangle$ develops long range order
($\langle\Phi\rangle \not = 0$). The ``Meissner phase" (or the $\Phi$-ordered
state) of (\ref{eiii})
corresponds to the state of the original GL theory (\ref{ei}) in which
only $N_{\Phi}$ field-induced vortex lines cross the system from one
end to another along ${\bf H}$. All other vortex excitations form either
closed thermally-generated loops of {\em finite} size or 
{\em finite} ``overhang" configurations decorating field-induced lines
as they make their way meandering 
from bottom to top of the sample. These field-induced vortex lines,
or s vortices in reformulation (\ref{eii}), have a 
finite line-tension
{\em relative} to the field direction and undergo effective ``diffusion"
along the z-axis (this is discussed in greater detail in the next section).
In the high temperature, ``normal metal" phase of (\ref{eiii}), the
$\Phi$-order is destroyed by expansion of thermally-generated vortex
loops and ``overhangs" decorating s (field-induced) vortices. We now
have new, thermally generated infinite loops ``percolating" all the way through the
system in all directions. These new infinite loops come on {\em top}
of the always present background of $N_{\Phi}$ s vortices. This is the
nature of the $\Phi$-transition\cite{zt} in the gauge theory (\ref{eiii})
and in the reformulation (\ref{eii}). The $\Phi$-transition is the {\em finite}
field version of the zero-field superconducting transition.\cite{zt} Its
thermodynamics, however, belongs to a {\em different} 
universality class: charged-superfluid (``inverted
3D XY") as opposed to neutral-superfluid (3D XY) at $H=0$, with finite $H$
playing the role of finite charge ($\tilde e^2\propto\sqrt{H}$) in the gauge
theory (\ref{eiii}).

The $\Phi$-transition line, $T_{\Phi}(H)$, 
plays a pivotal role in the gauge theory scenario. 
Since, on general grounds,\cite{physicac}  
we do not expect any true criticality associated with
the first-order vortex lattice melting line in 3D, $T_{\Phi}(H)$ is the {\em only}
critical line in the $H-T$ phase diagram of the original GL theory (\ref{ei}) and controls
fluctuation thermodynamics and transport at weak magnetic fields. It decides the
issues of relevance or irrelevance of various terms that can be added to 
(\ref{ei}) (point or columnar disorder, true electromagnetic screening with
finite $\kappa$, etc.) and provides the foundation on which one can build
a meaningful phenomenology of extreme type-II superconductors. In this respect,
$\Phi ({\bf r})$, the new ``superconducting" order parameter characterizing the
``line liquid" state, is ``more fundamental" than the original $\Psi ({\bf r})$.
This will now be amply illustrated.

To start building such phenomenology, we first need a reasonable
estimate of $T_{\Phi}(H)$ (Fig. 1). 
It starts at the zero-field superconducting transition $T_{c0}$, where
$\Phi ({\bf r})$ and $\Psi ({\bf r})$ are one and the same: the $H\to 0$
limit of (\ref{ei}) coincides with the $N_{\Phi}\to 0$ limit
in Eq. (\ref{eii}) and with the $\tilde e\to 0$ limit of the
gauge theory (\ref{eiii}). At finite, but weak field, we are in the
``extreme type-II" regime ($\kappa_s\gg 1$) of the gauge theory (\ref{eiii})
and we expect that the $\Phi$-transition is continuous and immediately becomes ``inverted".
The transition temperature $T_{\Phi}(H)$ is gradually reduced as function
of $H$ (or, equivalently, $\tilde e$ in (\ref{eiii})), due primarily to the reduction
in the effective line tension of very large ($\gg\ell$) vortex loops
caused by ``screening" generated by the ``medium" of 
field-induced vorticity. At these low fields,
$T_{\Phi}(H)$ can be evaluated directly from (\ref{eiii}). As $H$ increases,
however, numerous additional terms present in Eq. (\ref{eii}), but 
not included in the gauge theory (\ref{eiii})
on the grounds of their RG irrelevance at long distances ($\gg\ell$), 
start affecting $T_{\Phi}(H)$. Among
such terms none are more important than 
short distance ($\sim\ell$) positional correlations
which eventually lead to s vortex lattice formation at low temperatures. 

In general, the $\Phi$-transition and the vortex lattice melting 
are two completely different phase transitions, with two different order parameters, driven
by two different mechanisms. One is a $q\to 0$, another $q\sim 1/\ell$ transition.
They are not entirely unrelated, however, since they arise in the same theory,
(\ref{ei}) or (\ref{eii}). For instance, as $H\to 0$, we must have $T_{\Phi}(H) \geq
T_m(H)$.\cite{reversal} This is so because only in the $\Phi$-ordered state do s vortices in
(\ref{eii}) have finite long range interactions, 
$\propto \vert\langle\Phi\rangle\vert^2$.\cite{footv} Without such long range
interactions the s vortex system would remain in a liquid state as $H\to 0$.\cite{nelson}
Similarly, in the solid phase, s vortices form a lattice and cannot
screen large thermally-generated vortex loops, i.e. $\tilde e$ becomes effectively
zero even for $H\not =0$.
All vortex loops will then remain small and bound, just as they were at $H=0$.
This is discussed in more detail in Sec. VII. The problem is that the melting
transition is {\em always first-order} and thus, in principle, we could have
$T_{\Phi}(H) = T_m(H)$ over most of the $H-T$ phase diagram
or even for {\em all} $H$. This would
mean that melting is so strongly discontinuous that it always ``jumps"
over the intermediate, $\Phi$-ordered phase, straight into the
true normal state. I know of no argument to rule out this possibility. 

This being said, the most likely outcome
is the one depicted in Fig. 1. At higher fields, as we approach the ``type-I" regime
of (\ref{eiii}) ($H\sim H_s$ (\ref{ev})), the gauge theory suggests that $\Phi$-transition
itself {\em converts} to {\em first-order}. In this situation, 
it seems justified to assume
that $T_{\Phi}(H)=T_m(H)$, as shown in Fig. 1. For low fields, $H\ll H_s$, where
the melting transition becomes {\em weakly} first-order and
(\ref{eiii}) predicts a strong ``type-II" behavior and {\em continuous}
$\Phi$-transition, it is natural to expect $T_{\Phi}(H) > T_m(H)$. 
At fixed low field, as we increase the temperature in Fig. 1, {\em both} the
effective strength of the Biot-Savart interaction between s vortices (Sec. VII) and their
effective ``mass" (Sec. V) {\em decrease}. As interactions and line tension go down,
a natural progression of thermodynamic phases
follows: a solid (Abrikosov lattice), a ``massive" liquid ($\Phi$-ordered phase
or ``line liquid") and, finally, a ``massless" fluid of unbound loops (a true
normal state). A mean-field calculation, performed in Ref. 9, 
indeed leads to such results. I propose here a 
simple criterion which summarizes the results of such calculations and can be used
to determine $T_{\Phi}(H)$ and $T_m(H)$ at low fields,
$H\ll H_s$: the s vortex lattice melts when the average size of thermally-generated
loops, $\Lambda_{\Phi}(T,H)$, reaches a fraction $d_m$ of the average distance between
field-induced (s) vortices: $\Lambda_{\Phi}(T,H)=d_m\sqrt{2\pi}
(\xi_{\parallel}/\xi_{\perp})^{1/3}\ell$.  
$d_m\sim 0.2 - 0.3 $ and $\Lambda_{\Phi}(T,H)\approx\Lambda_{\Phi}(T,0)
\sim (\xi_{\perp}^2\xi_{\parallel})^{1/3}$ seems a reasonable estimate. 
Here $\xi_{\perp,\parallel} =
\xi_{0\perp,0\parallel}|t|^{-\nu}$ are the true superconducting correlation
lengths at $H=0$.
This results in the expression for the vortex lattice transition temperature in
the critical region:
\begin{equation}
t_m (h) = - \frac{1}{(d_m\sqrt{2\pi})^{3/2}}
\bigl(\frac{\xi_{0\perp}}{\xi_{GL,\perp}}\bigr)^{3/2}
h^{3/4} ~~~,
\label{eviii}
\end{equation}
where the temperature is measured relative to the {\em true} zero-field
superconducting transition, $t=(T/T_{c0}) -1 $, $h$ is defined below Eq. (\ref{evi})
and $\nu_{xy}$ was set to 2/3. 
The ratio $\xi_{0\perp}/\xi_{GL,\perp}$ should be $\sim 1$.
As argued above, we expect $d_m\sim 0.2-0.3$.
The $\Phi$-transition, on the other hand, takes place when the size
of thermally-generated loops, at {\em finite} $H$,  reaches the sample dimensions,
$\Lambda_{\Phi}(T,H)\to \infty$. This should take place along the line where
the average loop size for $H=0$, $\Lambda_{\Phi}(T,0)$, becomes of the
order of average distance between s vortices, i.e.
$\Lambda_{\Phi}(T,0) =d_{\Phi}\sqrt{2\pi}(\xi_{\parallel}/\xi_{\perp})^{1/3}\ell$, 
with $d_{\Phi} \sim 1$.  This
determines the vortex loop ``expansion" line, or $T_{\Phi}(H)$:
\begin{equation}
t_{\Phi} (h) = - \frac{1}{(d_{\Phi}\sqrt{2\pi})^{3/2}}
\bigl(\frac{\xi_{0\perp}}{\xi_{GL,\perp}}\bigr)^{3/2}
h^{3/4} ~~~,
\label{eix}
\end{equation}
with $d_{\Phi}\sim 1$.\cite{tphi} 
Obviously, $T_{\Phi}(H) > T_m(H)$ since $d_m < d_{\Phi}$. 
Eqs. (\ref{eviii},\ref{eix}) are valid only in the
limit of low fields, $H\ll H_s$ (\ref{ev}). At higher fields, $H\sim H_s$,
$T_{\Phi}(H)$ and $T_m(H)$ merge together and both vortex loop ``expansion"
and vortex lattice melting occur {\em simultaneously} when
$\Lambda_{\Phi}(T,H)$ reaches $\sim\sqrt{2\pi}\ell$ from within the solid phase
(Fig. 1). The above expressions (\ref{eviii},\ref{eix}),
with $d_m$ and $d_{\Phi}$ serving as numerical parameters, can be viewed as a 
``Lindemann criterion" for vortex loops and should provide reasonable
estimates of $T_{\Phi}(H)$ and $T_m(H)$.
The reader should bear in mind that the above 
expressions for $T_m(H)$ and $T_{\Phi}(H)$ are only estimates,
however reasonable. Only the first-principle 
calculation, starting from (\ref{ei}) or (\ref{eii}), 
and computing accurately free energies
of different phases (solid, $\Phi$-ordered liquid, and the true normal state) can
settle this question within the theoretical framework proposed in this paper. 
Unfortunately, such calculation is not feasible at the present level
of analytical sophistication. Additional careful
experiments on HTS and numerical simulations of 3D XY and related models
are far more promising in this regard.  

I now proceed to further investigate the phase diagram represented by Fig. 1 and
Eqs. (\ref{eviii},\ref{eix}).
As the field is turned on in Eq. (\ref{ei}) we can immediately write down the 
scaling expression
for the dimensionless singular part of the free energy, $f$,
associated with critical fluctuations:
\begin{equation}
f= \vert t\vert^{2-\alpha}\phi _{\pm}(\frac{H}{H_k|t|^{\Delta}})~~~,
\label{ex}
\end{equation}
where $t=(T/T_{c0}) -1$ and $H_k$ depends on material parameters.
This expression is completely general and as such conveys little information. It is
based only on the existence of the {\em zero-field} critical point. The same
expression can be written for spin systems or any other system exhibiting
a critical point which is then perturbed by a generalized ``field".\cite{goldenfeld}
In our case, the $H=0$ critical point is in the 3D XY universality class
and we should have $\alpha =\alpha_{xy}$. Furthermore, based on dimensional
analysis and general physical arguments, it was proposed in Ref.\cite{ffh} 
that $\Delta = 2\nu_{xy}$. This result holds to two-loop order in
RG\cite{lawrieii} and is likely exact, as emphasized in Sec. II.
$\phi _{\pm}(\frac{H}{H_k|t|^{\Delta}})$ is a universal function of
its argument inside the 3D XY critical region of the GL theory.
At present, its form is not known.
Note that Eq. (\ref{ex}), while completely general, is written in the
form which implicitly suggests that the finite-field critical behavior is governed
by the zero-field critical {\em point}, as is frequently the case
in spin systems.

In the gauge theory scenario, the situation is different and we can
be more specific. First, as already emphasized, within this
scenario the critical fluctuations are governed by a critical {\em line}
and not a critical {\em point}.\cite{zt} This means that we immediately learn something
about the function $\phi_{\pm}$
defined in Eq. (\ref{ex}): $\phi _{\pm}$ is {\em non-analytic} 
along the $\Phi$-transition line, $T_{\Phi}(H)$ (\ref{eix}).
This line singularity should be explicitly incorporated into the expression
for the free energy. To devise such a new scaling function, based on the
gauge theory scenario, we start by observing that we can eliminate
the ``trivial" part of the charge anisotropy from (\ref{eiii}) by
rescaling all lengths and fictitious vector potential ${\bf S}$ with
an appropriate superconducting correlation length, $\xi_{\perp,\parallel}$,
in the way that makes the $\Phi$-dependent part of (\ref{eiii})
isotropic. This rescaling procedure is a variation on the
familiar rescaling of anisotropy at the $H=0$ transition.
After the rescaling, the $\Phi$-dependent part of (\ref{eiii}) 
describes an isotropic superconductor with a correlation length
$\xi = (\xi_{\perp}^2\xi_{\parallel})^{1/3}$, while the coupling 
constants in the last two terms become:
\begin{equation}
K_{\perp,\parallel}\to 
K'_{\perp,\parallel}= c_{\perp,\parallel}(\xi_{\parallel}/\xi_{\perp})^{1/3}T\ell~~.
\label{exi}
\end{equation}
The following quantities appear in Eq. (\ref{exi}):
$\xi_{\perp,\parallel} = 
\xi_{0\perp,\parallel}|t|^{-\nu}$ and
$\xi = (\xi_{\perp}^2\xi_{\parallel})^{1/3}=
\xi_0|t|^{-\nu}$ are the true diverging superconducting
correlation lengths at $T_{c0}$, defined by the eigenvalues of the helicity modulus
tensor (see the Appendix B). 
Accordingly, $\Gamma = \xi_{\perp}/\xi_{\parallel}$ is 
the {\em true} anisotropy ratio at the $H=0$ critical
point ($T_{c0}$) and {\em not} the GL anisotropy,
$\Gamma_{GL} = \xi_{GL,\perp}/\xi_{GL,\parallel}$.
It now becomes clear why $K_{\perp,\parallel}$ have been defined in Eq. (\ref{eiii})
with the anisotropy $\Gamma$ explicitly factored out:
new rescaled coupling constants can simply be written as:
$K'_{\perp,\parallel}= c_{\perp,\parallel}T\bar{\ell}$, where
\begin{equation}
\bar{\ell} = \frac{\ell}{\Gamma ^{1/3}}~~{\rm corresponds~to}~~\bar{H} = \Gamma ^{2/3}H~~.
\label{exii}
\end{equation}
$\bar{H}$ is just the rescaled magnetic field appearing in the original GL theory (\ref{ei})
{\em after} the anisotropy at the $H=0$ critical point has been rescaled out.
Consequently, $c_{\perp,\parallel}$ describe the {\em fundamental} anisotropy
of the gauge theory (\ref{eiii}),  which is inherent to the $H\not =0$ problem
and is not associated with a ``trivial" anisotropy at $T_{c0}$.\cite{charge,gauge}
The corresponding
fictitious ``charges" associated with $K'_{\perp,\parallel}$ are:
\begin{equation}
\tilde e_{\perp,\parallel}^2(T,H) = \frac{1}{c_{\perp,\parallel}(T,H)\bar{\ell}}~~~.
\label{exiii}
\end{equation}

The product of the above rescaling procedure is a fictitious anisotropic
electrodynamics with two ``charges", $\tilde e_{\perp}$ and $\tilde e_{\parallel}$.
As discussed in Sec. II, the charge is a relevant operator at the $H=0$
($\tilde e = 0$) critical point, with scaling dimension equal to 1/2.
We thus define two dimensionless scaling variables:
\begin{equation}
\tilde e_{\perp,\parallel}^2\xi = \frac{\xi}{c_{\perp,\parallel}\bar{\ell}} =
\sqrt{\frac{2\pi}{\phi _0}}\frac{\xi_0\Gamma ^{1/3}}{c_{\perp,\parallel}}
\frac{H^{1/2}}{|t|^{\nu}}~~.
\label{exiv}
\end{equation}
Note that dimensionless ratio $\xi/\bar{\ell}$ is common to both charges
and (non-trivial) anisotropy is stored in $c_{\perp}$ and
$c_{\parallel}$. $c_{\perp}$ and
$c_{\parallel}$, however, are also functions of $\xi/\bar{\ell}$ {\em only}. 
Therefore, there is only a {\em single} relevant 
scaling variable, the dimensionless charge:
\begin{equation}
q^2_0 = \frac{\xi}{\bar{\ell}} =
\sqrt{\frac{2\pi}{\phi _0}}\xi_0\Gamma ^{1/3}
\frac{H^{1/2}}{|t|^{\nu}}~~,
\label{exv}
\end{equation}
which is precisely the original scaling variable\cite{ffh} of the GL theory (\ref{ei}),
since $\xi/\bar{\ell} = \xi_{\perp}/\ell$.
The functions $c_{\perp,\parallel} (\xi/\bar{\ell})$ are discussed further
in the next section and Appendix B.

We are now in position to write down the scaling function for the free energy within the
gauge theory scenario, with the non-analytic part associated
with the $\Phi$-transition explicitly factored out:
\begin{equation}
f_s = \vert t - t_{\Phi}(h)\vert ^{2-\alpha}
\Omega_{\pm}^{L,S}~\Bigl(\frac{t-t_{\Phi}(h)}{\vert t_{\Phi}(h)\vert}\Bigr)~~~.
\label{exvi}
\end{equation}
Note that $t_{\Phi}(h)$, defined in Eq. (\ref{eix}), also follows 
from $q_0^2 = \sqrt{2\pi} d_{\Phi}$. $d_{\Phi}$ is therefore a universal
number of the GL theory (\ref{ei}), as is $d_m$.\cite{meltingline}
$\Omega _{\pm}^{L,S} (x)$ is a universal and
{\em regular} function of its argument. The subscripts
$\pm$ refer to $x >0$ ($x<0$), while the superscripts $L$ and $S$ indicate the
``vortex liquid" and ``vortex solid" branches of $\Omega$, respectively. For example,
below $T_m(H)$, we should use $\Omega _{-}^S (x)$.
In writing down (\ref{exvi}) I have assumed that the correlation length
exponent of the gauge theory $\nu_{GT}\sim\nu_{xy}\sim 2/3$ and that
the hyperscaling relation holds, resulting in $\alpha\sim\alpha_{xy}$.

How do we evaluate the crossover function, $\Omega (x)$?
I alert the reader to the following important point: the gauge theory scenario
explored in this paper allows one to, {\em in principle}, determine all the 
branches of the crossover function $\Omega (x)$ by using 
a combination of perturbation theory and RG techniques. 
Such analytic calculation is extremely laborious 
and far beyond the scope of this paper.  
Well informed reader will immediately realize that many aspects
of this calculation are computationally 
extremely demanding, and actually have {\em not} been accomplished in
the published literature even for the
ordinary $H=0$ situation. Indeed, the technical difficulties involved are 
of the same general nature. 
This, however, does not detract from
the main message of this section: the underlying physical picture
of the gauge theory scenario provides a systematic,
conceptually straightforward way to compute the $H\not = 0$
3D XY critical thermodynamics at the {\em same level} 
of analytical accuracy as is presently feasible for the $H=0$ case.

Faced with such odds, I assume, for the purposes of this paper, that 
$\Omega (x)$ in Eq. (\ref{exvi}) is some unknown universal crossover
function, to be determined either from numerical simulations\cite{sudbo} or directly
from experiments.  With the free energy thus specified, we can proceed to evaluate 
the {\em singular} part of all thermodynamic functions, simply by taking
requisite derivatives.\cite{friesen}

\section{Helicity modulus, line ``diffusion", topological windings, and 
physical nature of $\Phi$-order}
I now turn to physical properties which allow more direct look at actual
configurations of loops and lines
that characterize the state of a superconductor above and
below $T_{\Phi}(H)$. A useful measure of a degree of superconducting order
is a ${\bf q}$-dependent helicity modulus tensor, $\Upsilon ({\bf q})$, whose
components are defined as\cite{teitel} 
\begin{equation}
\Upsilon_{\mu\nu} ({\bf q}) =V\frac{\delta^2 F}{\delta {\bf a}_{\nu}({\bf q})
\delta {\bf a}_{\mu} (-{\bf q})}~~~, 
\label{exvii}
\end{equation}
where $\mu, \nu = x,y,z$, 
$V$ is the total volume, $F$ is the free energy of the GL theory (\ref{ei})
and ${\bf a}({\bf r})$ is a small (infinitesimal) vector potential
added to the external ${\bf A}$. The uniform component of the associated
magnetic field, ${\bf h}({\bf r}) = \nabla\times {\bf a}$, vanishes.
The above second derivative is evaluated in the ${\bf a}\to 0$ limit.

$\Upsilon ({\bf q})$ measures the ability of a system to ``screen" out tiny
external fields. In the {\em superconducting} phase $\lim_{{\bf q}\to 0}
\Upsilon ({\bf q})$ is finite and the system is said to exhibit a
differential Meissner effect. In a {\em normal} phase, $\Upsilon ({\bf q})
\sim q^2$ and vanishes in the $q\to 0$ limit. Within our ``helium model",
the way $\Upsilon$ is reduced to zero in the long wavelength limit is
through proliferation of {\em infinite} vortex loops/lines which go all
the way across the system and can act as ``free charges", screening
a weak external perturbation. In this intuitive sense, we can
think of a normal state as a vortex ``metal", while the superconducting
state is a vortex ``dielectric", with only vortex loops of {\em finite}
size present as thermal excitations. 

To compute the helicity modulus of our original GL theory 
one adds an infinitesimal ${\bf a}_{\mu}$ to ${\bf A}_{\mu}$ in
Eq. (\ref{ei}). If we now go to the reformulation (\ref{eii}) and
finally, through the coarse-graining procedure of Appendix A, end up with
our fictitious gauge theory, ${\bf a}_{\mu}$ appears as a small
addition to the ``vector potential" ${\bf S}_{\mu}$ in the
second (gradient) term of Eq. (\ref{eiii}). This implies that the
long wavelength ($q\ll 1/\ell$) form of the helicity modulus of the
gauge theory (\ref{eiii}) coincides with that of the original GL
theory (\ref{exvii}).  Using the gauge theory (\ref{eiii}) and
ignoring the anisotropy, we obtain that {\em below}
$T_{\Phi}(H)$ (see Appendix B for details):
\begin{equation}
\Upsilon ({\bf q}) = Kq^2 - \frac{K^2q^4}{T}\langle {\bf S}({\bf q})\cdot
{\bf S}({\bf -q})\rangle =
Kq^2 + {\cal O}(q^4)~~~.
\label{exviii}
\end{equation}
In the $\Phi$-ordered state our fictitious gauge field is ``massive", i.e.
exhibits a Meissner effect, and 
$\lim _{q\to 0}\langle {\bf S}({\bf q})\cdot
{\bf S}({\bf -q})\rangle \propto |\langle\Phi\rangle |^{-2}$ goes
to a {\em finite} value. The simple physics behind this is that
thermally-generated vortex loops in $\Phi ({\bf r})$ have average size that is
{\em finite} and do not contribute at all to $\Upsilon ({\bf q})$ in the 
$q\to 0$ limit. Furthermore, Eq. (\ref{exviii}) tells us that,
if $K_{\perp,\parallel}$ are {\em finite}, the $\Phi$-ordered phase
is {\em not} a superconductor and has a {\em finite} superconducting 
correlation length, both perpendicular and parallel to the external field
(see Appendix B for details),
\begin{equation}
\xi_{\parallel}\sim K_{\perp}/T\sim c_{\perp}\ell~~,~
\xi_{\perp}^2/\xi_{\parallel}\sim K_{\parallel}/T\sim c_{\parallel}\ell~.
\label{exix}
\end{equation}
This result is easily understood: with $H$ finite, there
now must be $N_{\Phi}$ field-induced
vortex lines moving about in the sample. Unless these (s) vortices are pinned
down, as it happens in the vortex-solid phase, they will be available to ``screen"
weak (infinitesimal) external fields and the system is always a vortex ``metal"
with the ``screening length" $K/T$.
In particular, according to our assumption ii), the system of field-induced
vortex lines also contains windings in the xy plane and such ``screening", although
anisotropic, is finite in all directions. 
Below $T_{\Phi}(H)$, where s vortices are exclusively
responsible for the vanishing $\lim _{q\to 0}\Upsilon _{\mu}({\bf q})$,
$c_{\perp,\parallel} (q^2_0)$ determine
the $\perp$ and $\parallel$ ``screening lengths" in the s vortex system, in units
of magnetic length $\ell$ (Eq. (\ref{eaix}) in Appendix B). 

Above $T_{\Phi}(H)$ the situation changes and thermally generated vortex
loops ``expand" across the system. Obviously, the helicity modulus 
still vanishes, but now there is an abrupt 
drop in the coefficient of the $q^2$ term:
\begin{equation}
\Upsilon ({\bf q}) = \bigl(K - G\frac{K^2}{T\xi_{\Phi}}\bigr)~q^2 =
K\bigl(1 - C\frac{\tau ^{\nu_{GT}}}{\sqrt{H}}\bigr)q^2~~,
\label{exx}
\end{equation}
where $\tau (T,H)=(T-T_{\Phi}(H))/T_{c0}$ and $C$ and $G$ are constants.
$\nu_{GT}$ is the thermodynamic exponent of the Meissner transition in our
fictitious electrodynamics (\ref{eiii}), and $\nu_{GT}\sim\nu_{xy}\sim 2/3$, as
argued in Sec. IV. The second term in the above equation arises from
$\lim _{q\to 0}\langle {\bf S}({\bf q})\cdot
{\bf S}({\bf -q})\rangle = G/\xi_{\Phi}q^2$ in Eq. (\ref{exviii}), 
right above $T_{\Phi}(H)$.  This implies that the
original superconducting correlations, measured by 
$\langle\Psi (0)\Psi ^*({\bf r})\rangle$, remain finite in all directions on
{\em both} sides of $T_{\Phi}(H)$, but
there is a {\em non-analytic} drop in the superconducting correlation length
at the $\Phi$-transition,  
as thermally generated loops proliferate through the system and additional
infinite vortices become ``free charges" and available to screen. The new
order parameter, $\Phi ({\bf r})$, however, {\em does} attain a true long range
order below $T_{\Phi}(H)$, i.e. $\xi_{\Phi}\to\infty$ as $T\to T_{\Phi}(H)$
from above. It is unfortunately rather difficult to measure
the $\Phi$-correlations directly, by probing some suitably defined ``helicity
modulus" associated with the $\Phi$-order. This would require defining quantities
which are configuration-dependent and highly non-local, a rather
time-consuming proposition in a typical numerical simulation of a 3D XY
or related model. 

Still, the situation is far from hopeless.
We can devise another set of criteria  that are relatively easy to implement
in numerical simulations and yet
allow for a rather intimate look at the $\Phi$-order and
what precisely takes place as
we cross the $\Phi$-transition line. Below $T_{\Phi}(H)$, thermally generated
vortex loops are bound and field-induced vortices execute an effective
``diffusive" motion along the field direction. An average transverse displacement 
of a single field-induced vortex line from the point where it starts at $z=0$
to its ending point at $z=L_z$ goes as 
\begin{equation}
\sqrt{\langle r_{\perp}^2\rangle}
\sim \sqrt{D_s}L_z^{p}~~~,~~~p\cong \frac{1}{2}~~,
\label{edi}
\end{equation}
where $D_s$ is the effective ``diffusion" constant. This is shown 
in Fig. 3. The cutting/reconnecting of vortex lines does not affect this
diffusion process except by renormalizing $D_s$, as long as we are in the
$\Phi$-ordered phase.\cite{diffusion} For example, in the 3D XY model, where the
identification of an individual field-induced line is not unique, we should simply
look at the {\em distribution} of distinguishable vortex paths, obtained
by {\em randomly} resolving all the crossings, and
average over all distinct configurations. Such distribution will be ``diffusive",
with the average rms displacement given by Eq. (\ref{edi}).
We can use this effective ``diffusion" constant $D_s$ (\ref{edi}) to 
define an {\em effective} ``mass" in 
the elegant non-relativistic boson analogy of Nelson\cite{nelson}. Note
that the worldlines of such flux-bosons do {\em not} correspond to (s)
vortex lines in individual configurations of an extreme type-II 
superconductor. This is clear since non-relativistic bosons describe
{\em strictly directed} lines, i.e. contain no ``overhang" configurations
as they advance from bottom to top of the system along z-axis (the ``time"
axis in the boson analogy).
Such ``overhang" configurations, plus numerous vortex loop
excitations floating around, describe worldlines of ``particle-antiparticle"
creation processes and cannot be accommodated within the non-relativistic
quantum boson analogy. Still, as long as we are in the $\Phi$-ordered phase,
it is only the $N_{\Phi}$ field-induced vortex lines that go all the way
across the system. We can then define an {\em effective} system of
$N_{\Phi}$ flux-bosons in the boson analogy, with suitably adjusted bare mass
$m_s$ and effective interactions, so that its {\em long distance} ($\gg\ell$ and
$\gg \Lambda_{\Phi}(T,H)$) behavior 
faithfully emulates an extreme type-II superconductor (Appendix A).
Above $T_{\Phi}(H)$, as infinite
tangles of field-induced and thermally-generated vortices proliferate across
the sample in {\em all} directions, $D_s\to\infty$ ($m_s\to 0$)
and we get {\em hyperdiffusion}:
\begin{equation}
\sqrt{\langle r_{\perp}^2\rangle}\sim L_z^{p'}~~~,~~p'\sim 1~~.
\label{edii}
\end{equation}
This hyperdiffusion arises through processes depicted in Fig. 3,
where a vortex line winding along the field direction {\em simultaneously}
winds all the way in the xy-plane by 
connecting itself to thermally-generated tangles, which are
naturally present in the $\Phi$-disordered phase. In the 3D XY model, this
implies that the distribution of transverse 
displacements of individual field-induced vortex paths 
is no longer ``diffusive" and has rms displacement 
limited only by the system size (\ref{edii});
more precisely, the distribution of $r_{\perp}^2$ acquires a power-law
tail above $T_{\Phi}(H)$.
Such windings in the xy-plane are plainly in evidence in the recent numerical simulations
of Nguyen and Sudb{\o}\cite{sudbo}, right above their melting line.
With such additional xy windings present with a finite weight
in the partition function, the ``effective mass" $m_s$
of non-relativistic flux-bosons {\em vanishes} since the vortex line tension
{\em relative} to the field direction has gone to zero. 
Above $T_{\phi}(H)$, infinite vortex paths of length
$\sim L^2$ (assuming $L_{\perp}=L_{\parallel}=L$ and a simple
random walk) crossing the system in all directions 
contain a {\em finite} fraction of all vortex segments: these
are the ``massless excitations". 
In this respect, the $\Phi$-transition corresponds 
to the restoration of ``relativistic
invariance" in a dual system of quantum particles whose worldlines
are our original vortex loops and lines. 
To wit, the ground state 
of such quantum system,  containing only 
``vortex matter"\cite{crabtree} below $T_{\Phi}(H)$, explodes
with ``vortex matter", ``vortex antimatter" and 
``vortex tachyons"\cite{tachyons} (Fig. 3), as the ``vacuum" becomes
unstable at $T_{\Phi}(H)$ to spontaneous creation of ``particles" (Appendix A).

The above connection between the ``$\Phi$-order" and 
the effective line tension of field-induced
(s) vortex lines reveals directly the physical content
of the gauge theory (\ref{eiii}) and permits us to construct a purely
geometrical picture of the $\Phi$-transition.
To do so, consider once again
the normal state of an extreme type-II superconductor
or a 3D XY model at $H=0$. 
Above $T_{c0}$, we can find vortex paths
that go all the way from one end of a sample to another, in any direction (Fig. 2).
If we work with periodic boundary conditions in all directions, this statement
means that we have some windings along x-, y-, and z-axes. To understand what
is precisely meant by such windings, we wrap our system on a three-dimensional
generalized torus, embedded in a four-dimensional space--This is just
a geometrical way of representing periodic boundary conditions. Now, the
number of windings along, say,  x-axis, ${\cal N}_x$, is the {\em total} number
of continuous vortex paths in the whole system that wind all
around the torus in the x-direction, {\em irrespective} of their
orientation, i.e., whether they are ``vortex" or ``antivortex" paths relative
to the x-axis. Such paths are topologically distinct from finite closed vortex
loops: the latter can be continuously shrunk to a point while the
former can not. ${\cal N}_x$ is different from the winding number, $W_x$:
$W_x$ also counts all the windings along x-axis but with a single ``vortex"
path contributing $+1$ while an ``antivortex" path counts as $-1$. 
In the widely recognized vocabulary of the 2D XY model, ${\cal N}_x$ would correspond 
to the total number of vortices {\em plus} antivortices in the yz-plane,
while $W_x$ would be the total number of vortices {\em minus}
the total number of antivortices. Back in 3D, in the superconducting state below $T_{c0}$,
all thermally-generated vortices come in the form of finite closed loops and
both ${\cal N}_x = 0$ and $W_x =0$. Above $T_{c0}$, $W_x$ must
remain equal to zero due to the ``vortex neutrality" of the GL theory (\ref{ei})
or the 3D XY model, but ${\cal N}_x$ is now finite
and ${\cal N}_x\propto L^{1+u}$, where $L$ is the linear size of the system
(we are assuming $L_{\perp}=L_{\parallel}=L$) and $u$ is the
``anomalous dimension" of such infinite paths. This implies that there is a {\em finite}
fraction of all vortex segments contributing to such windings along x in the yz-plane. 
The same holds for the winding number and the total number of
windings along y- and z-axes, $W_{y(z)}=0$ and ${\cal N}_{y(z)}\propto L^{1+u}$.
This can be summarized as:
$${\cal N}_{x,y,z}=0~~{\rm for}~T<T_{c0}~,$$
\begin{equation}
{\cal N}_{x,y,z}= 2n^{T}_{x,y,z}L^{1+u}~~{\rm for}~T>T_{c0}~,
\label{exxi}
\end{equation}
where $n^{T}_x$ is the ``density" in the yz-plane
of thermally-generated infinite vortex-antivortex winding
paths traversing the system along the
x-axis, and so on. In the isotropic case $n^{T}_x =n^{T}_y =n^{T}_z$. Of course,
the presence of such windings in all directions is the reason why the material
is not in the superconducting state above $T_{c0}$: these infinite vortex paths
can now move to ``screen" weak external fields, driving the helicity modulus to zero
in the long wavelength limit and producing finite dissipation.

As we turn on a finite field in (\ref{ei}), we still have
$W_{x(y)}=0$ but $W_z = N_{\Phi}$ and consequently ${\cal N}_z$ must be at least $N_{\Phi}$
in every configuration of the system.\cite{zt} Imagine now how the state of the
system evolves along a
small circle in the $H-T$ phase diagram (Fig. 1), surrounding $T_{c0}$.
Our circular path starts at $H=0$ and at some temperature $T$ slightly {\em above} $T_{c0}$,
and evolves in the counter clockwise direction toward its end point at
$H=0$ and some temperature slightly {\em below} $T_{c0}$. 
Initially, we are 
very close to the $H=0$ normal state and it is safe to assume that
$N_{\Phi}L\ll {\cal N}_z (T,H=0)L^{2-u}$ at some temperature $T$ slightly above $T_{c0}$. 
In this case
it is natural to expect that, in a finite field:
\begin{equation}
{\cal N}_{x,y}= 2n^{T}_{x,y}L^{1+u_{\perp}}~~~,
~~~{\cal N}_z= n_{\Phi}L^2 + 2n^{T}_zL^{1+u_{\parallel}}~~~,
\label{exxii}
\end{equation}
where $n_{\Phi}=N_{\Phi}/L^2=1/2\pi\ell^2$ is the density of field-induced
(s) vortex lines and $n^{T}_{x,y,z}(T,H)$ are 
``densities" of thermally-generated vortex-antivortex windings.
Note that now $2n^{T}_{x,y}\not = n_{\Phi} + 2n^{T}_z$
and $n^{T}_{x,y}\not =n^{T}_z$
even in the isotropic case (although the gauge theory scenario strongly
suggests $u_{\perp}=u_{\parallel}$).
There is a finite ``density" of thermally generated infinite vortex (or antivortex)
winding paths in any direction. Eq. (\ref{exxii}) describes how the
normal state of the system (\ref{exxi}) has changed after the application of
a finite, but weak external field.

On the opposite side of our imaginary cirle, near its end point at $H=0$ and $T<T_{c0}$,
the situation is completely different. Now, the zero-field state is a
{\em superconductor} and ${\cal N}_{x,y,z}=0$. Any finite field, no matter
how small, has a drastic effect.  For very low fields,
it is natural to expect that there are no thermally-generated
infinite vortex loops and only those windings associated
with the field-induced vortex lines are present in the system:
\begin{equation}
{\cal N}_{x,y}= 0~~~,
~~~{\cal N}_z= N_{\Phi}=n_{\Phi}L^2~~~.
\label{exxiii}
\end{equation}
This is just the (s) vortex lattice state in Fig. 1. 
Note that in these general geometrical terms there 
is no difference between the (s) vortex lattice
state and the ``anisotropic superconducting liquid" of 
Feigel'man et al.\cite{nordborg}.
Due to the absence of windings in the xy-plane both have superconducting
response along the field direction, i.e. the $\Upsilon _{zz} ({\bf q})$
component is finite in the $q\to 0$ limit. Also, in both
cases,  $\Upsilon _{xx(yy)}({\bf q})$ vanish as $q\to 0$.
The only difference is that it takes arbitrary weak pinning to restore
superconductivity in all directions in the vortex lattice state.
On this basis, I have assumed in the phase diagram of Fig. 1 that 
such an ``anisotropic superconducting liquid" phase is preempted
by the first-order transition at $T_m(H)$.\cite{feigelman}

In our proposed phase diagram depicted in Fig. 1, the intermediate
$\Phi$-ordered phase is inserted between the true normal state (\ref{exxii}) and 
the vortex lattice phase (\ref{exxiii}). What is its nature in simple 
geometrical terms used to describe the other two phases in Eqs. (\ref{exxii})
and (\ref{exxiii})? In the $\Phi$-ordered state all thermally-generated
vortex loops are bound and only $N_{\Phi}$ field-induced (s) vortex lines
go from one end of the sample to another, along ${\bf H}$, resulting
in precisely $N_{\Phi}$ windings along the z-axis. However,
these (s) vortex lines are in a liquid state, characterized by some
finite effective line tension, ${\cal T}$, and they ``diffuse"
in the xy-plane while winding along ${\bf H}$.
This translates into a {\em non-vanishing} total number of windings in the
xy-plane:
\begin{equation}
{\cal N}_{x,y}= {\cal I}_{x,y}L^{2p}~~~,
~~~{\cal N}_z= N_{\Phi}=n_{\Phi}L^2~~~.
\label{exxiv}
\end{equation}
We expect $1>p\cong 1/2$.\cite{diffusion}
${\cal I}_{x,y}$ are some finite quantities having 
dimension of (length)$^{-2p}$
and we use Eq. (\ref{exxiv}) as their definition.
Eq. (\ref{exxiv}) describes the $\Phi$-ordered state using a simple geometrical
language of this section. 
The fact that ${\cal N}_{x,y}\not =0$ has nothing to do with the
thermal ``expansion" of vortex loops; these are still all of finite
size. Rather, it is due to the lateral ``diffusion" of field-induced
lines, as they wind along the z-axis. The field-induced vortices tend to form
infinite ``clusters", which manage to wind a {\em finite} number of times,
${\cal N}_{x,y}^c$, along x(y) by winding {\em infinitely} many times,
${\cal N}_{z}^c \sim L$, along the field direction (I now set $p=1/2$).
For such infinite clusters, we can {\em define} a long
distance, effective ``diffusion" constant, ${\cal D}$, 
and the associated line tension, ${\cal T}$, both relative to the field direction:
\begin{equation}
{\cal D}_{x,y} = 
\frac{{\cal N}_{x,y}^c L}{{\cal N}_z^c}~~,~
{\cal T}_{x,y} \sim {\cal D}_{x,y}^{-1}~~.
\label{exxv}
\end{equation}
Defined in this fashion, ${\cal D}$ and ${\cal T}$ are truly global
quantities, detached from complicated individual configurations of
interacting vortex loops and lines and
dependent only on thermodynamic state of the system as a whole.
In the $\Phi$-ordered state both are finite and have a well defined
average. In the true normal state, however, the infinite clusters
appear for which ${\cal D}\to \infty$ and ${\cal T}\to 0$. 
Infinite vortex paths inside such 
clusters manage to wind in the x(y) direction after only a 
{\em finite} number of windings along the field.
In this sense, ${\cal T}$ can serve as a probe for the
presence/absence of the $\Phi$-order. Above $T_{\Phi}(H)$,
such clusters with ${\cal T} = 0$ contain a {\em finite} fraction of
all vortex segments.
Again, in the $\Phi$-ordered state of
the 3D XY model, where there is no unique identification of individual vortex paths
due to their crossings, every configuration that contributes to the
thermodynamic limit has precisely 
$N_{\Phi}$ vortex lines going from bottom to top 
in every distiguishable assignment of such paths. All other vortex paths
form finite closed loops. The {\em distribution} of the total number of 
windings, obtained by randomly resolving all the crossings,
still satisfies (\ref{exxiv}).

Once the system makes this phase transition to the normal state (\ref{exxii})
and the $\Phi$ order is lost, there is only a smooth evolution.
The ``densities" of thermally-generated windings $n^{T}_{x,y,z}$,
as well as the way these windings are
realized in individual configurations of vortex
loops and lines, can change considerably, depending on where
we are in the $H-T$ phase diagram relative to $T_{\Phi}(H)$,
but we do not expect to cross any additional phase boundaries.

The geometrical picture presented here, based only on global topological
properties of loops and lines, gives a clear insight into two forms
of ``superconducting" order, described by two {\em different} order parameters,
$\Psi ({\bf r})$ and $\Phi ({\bf r})$. The familiar superconducting
order, measured by $\Psi ({\bf r})$, reflects 
the system's ability to expel tiny external fields and 
is manifested by a finite helicity modulus and absence of linear dissipation.
This leads to the well-known spectacular experimental consequences and is naturally
of great practical importance. A more subtle form of ``superconducting" order, 
associated with the new order parameter $\Phi ({\bf r})$ introduced in Ref.\cite{zt},
describes the presence of finite line tension at all lengthscales and is manifested by 
suppression of large thermally-generated vortex loops in the partition
function. At $H=0$, or in an infinitesimal field with only a single
field-induced line, $\Psi$ and $\Phi$ are equivalent. The  helicity modulus and 
the effective line tension vanish simultaneously at 
a {\em single} superconducting transition, $T=T_{c0}$. In a {\em finite} field,
with a finite density of field-induced lines,
the situation is different: the ``superconducting" transition can now be viewed as 
{\em split} in two branches, $T_m(H)$ and $T_{\Phi}(H)$. At $T_m(H)$ the standard
superconducting order is lost as the vortex lattice melts into a 
liquid of field-induced (s) vortex lines. Even though the total number
of windings along ${\bf H}$ is still locked at $N_{\Phi}$, just like
in the vortex lattice state, the effective ``diffusion" of (s) vortex
lines leads to windings in the xy-plane, ${\cal N}_{x,y}\propto L^{2p}$.
This amount of winding {\em suffices} to 
cause vanishing of the helicity modulus (\ref{exviii})
and drives $\Psi ({\bf r})$ to zero (i.e., $\langle\Psi (0)\Psi^* ({\bf r})\rangle$
is short-ranged). 
In contrast, the long distance line tension 
${\cal T}\propto 1/{\cal D}$ (\ref{exxv}) is still finite below $T_{\Phi}(H)$,
same as in the Meissner state of the $H=0$ superconductor. 
This state {\em differs} from the normal state by the presence of
long range $\Phi$-order, measured by
$\langle\Phi (0)\Phi^* ({\bf r})\rangle$. 
It can be considered {\em isomorphic}, in the {\em long wavelength
limit}, to a liquid (or solid, as shown in Fig. 1) 
of $N_{\Phi}$ field-induced vortex lines with some {\em finite}
effective line tension relative to the field direction, and interacting
via a long range, London-Biot-Savart-type interactions, whose
overall strength is set by $\sim\vert\langle\Phi ({\bf r})\rangle\vert^2$.\cite{footv}
Only at some higher temperature $T_{\Phi}(H)$,
does ${\cal T}$ vanish and infinite vortex loops proliferate across the system.
This signals the destruction of the $\Phi$-order as the system finally makes transition
to the true normal state. 
Above $T_{\Phi}(H)$, {\em both} $\Psi$ and $\Phi$-order are absent. Consequently,
the {\em true normal} state of
the GL theory in a finite magnetic field (\ref{ei}) should not be viewed
as a ``line-liquid" in the commonly accepted sense\cite{blatterrmp}. 
The general geometrical arguments of 
this section preclude such identification and
point to a direct connection between the
presence (absence) of the $\Phi$-order and our ability (inability) to
emulate the long wavelength behavior of the system (\ref{ei}) in
terms of a conventional ``line-liquid" (or a ``line-solid") of field-induced
vortices. In the language of the {\em non-relativistic} ``boson 
analogy", the mass of the bosons {\em vanishes}
at $T_{\Phi}(H)$ and such analogy breaks down in the true normal state.

The above criteria should allow one
to clearly distinguish the high and the low temperature states
of the system in numerical simulations
and to establish whether they are separated by a true
thermodynamic phase transition or a sharp crossover. 
Such procedure should be superior to measurements of specific heat which
possesses at most nearly-logarithmic singularity and is severely
limited by the finite size effects.

\section{Fluctuation conductivity}
Electrical conductivity (or resistivity), is a quantity easily 
measured experimentally. Unfortunately, fluctuation conductivity, while in principle
a very useful probe of a degree of superconducting order, is not purely
thermodynamic quantity and cannot be evaluated from the GL theory (\ref{ei})
unless additional assumptions are made concerning the time-dependence of
the fluctuating superconducting order parameter. We adopt here a 
frequently used and empirically successful assumption of dynamical scaling,\cite{ffh}
which connects the decay of spatial correlations with that of time 
correlations: $\tau _{\rm sc}\sim \xi_{\rm sc}^z$, where 
$\tau _{\rm sc}$ is the relaxation time associated
with the superconducting order parameter and $z$ is the dynamical
critical exponent. At zero field, this assumption leads to the expression for
the fluctuation conductivity:
\begin{equation}
\sigma\propto \xi_{\rm sc}^{z + 2 - D}\sim t^{\nu_{xy} (D-2-z)}~~~,
\label{exxvi}
\end{equation}
with $z\sim 1.5$ from numerical simulations.\cite{girvin}
For simplicity, we have suppressed anisotropy in the above expression. When
a finite external field is turned on, and we are 
inside the critical region near
$T_{c0}$, where $\xi_{\rm sc}$ is very long, we can still write:
\begin{equation}
\sigma\propto \bigl[\xi_{\rm sc}(H\not =0)\bigr]^{z + 2 - D}~~~.
\label{exxvii}
\end{equation}
Strictly speaking, the dynamical exponent could be different in the finite field
case but I will ignore such possibility. I also
concentrate on dissipative transport and do not consider
Hall conductivity. Right below $T_{\Phi}(H)$, the gauge
theory (\ref{eiii}) suggests that $\xi_{\rm sc}$ is finite in all directions
and $\xi_{\rm sc}\propto K/T\sim c\ell$, as discussed in the Appendix B.  
Above $T_{\Phi}(H)$, the screening length $\Lambda$, defined
in the Appendix B by Eq. (\ref{eaix}),
drops abruptly (\ref{exix}) and we {\em assume} that the 
superconducting correlation length $\xi_{\rm sc}$ does the same,
\begin{equation}
\xi_{\rm sc}\propto c\ell (1 - G\frac{c\ell}{\xi_{\Phi}})
= c\ell(1 - C\frac{\tau ^{\nu_{GT}}}{\sqrt{H}})~~~.
\label{exxviii}
\end{equation}
This assumption seems perfectly justified on physical grounds, since
the drop in the screening length arises through appearance above
$T_{\Phi}(H)$ of thermally-generated infinite vortex loops which lead to
additional screening. It is natural to expect that these same loops
suddenly increase dissipation and produce a non-analytic drop in the conductivity.

After restoring the anisotropy, the relative change in the conductivity,
$\delta\sigma _{\mu}$,
as one crosses over from the $\Phi$-ordered to the true normal state is:
\begin{equation}
\delta\sigma _{\mu} =\frac{\sigma_{\mu,<} -\sigma_{\mu,>}}{\sigma_{\mu,<}}=
(z+2-D)C_{\mu} \frac{\tau ^{\nu_{GT}}}{\sqrt{H}}~~~,
\label{exxix}
\end{equation}
where $\mu = (\perp, \parallel )$,
$\sigma_{\mu,<,>}$ is the fluctuation conductivity below (above)
$T_{\Phi}(H)$, $C_{\mu}$ are constants depending on material parameters,
and $\nu_{GT}\sim\nu_{xy}\sim 2/3$. The vortex loop ``expansion" that
takes place at $T_{\Phi}$ leads to a non-analytic increase in dissipation
and a corresponding drop in conductivity. 

\section{Vortex lattice melting in the critical region}
A theory of the vortex lattice melting in the critical
region is an elaborate subject and its detailed discussion
will be presented elsewhere. There are, however, several important
consequences of the present gauge theory scenario that concern the
very nature of the melting transition. We therefore discuss here
the ``minimal" set of requirements that should be satisfied by
any theory of melting consistent with the gauge theory scenario.
We should first observe that
the reformulation (\ref{eii}) obviously has a 
(s) vortex lattice as its ground state at low temperatures, just
as the original GL theory (\ref{ei}). The gauge theory (\ref{eiii}), however,
does {\em not}, for the simple reason: up to this point we were interested in the
{\em long wavelength}, $q\ll 1/\ell$, behavior. The $\Phi$-transition, for example,
is clearly the $q\to 0$ transition. On this basis, we have dropped a 
large number of terms from Eq. (\ref{eiii}), by arguing that they are irrelevant
at very long distances. The melting transition, in contrast, is
a {\em finite} $q$-transition ($q\sim 1/\ell$), and it requires additional
terms and modifications to the coarse-graining procedure applied on the
way from (\ref{eii}) to (\ref{eiii}). For instance, higher powers
of ${\bf S}$, particularly the odd ones, reflecting the up-down asymmetry
along ${\bf H}$ manifest in Eqs. (\ref{ei}) and (\ref{eii}), where unimportant
in the long wavelength limit but are {\em essential} for the transition
to a non-uniform (s) vortex lattice state. 

This being so, the ${\Phi}$-transition
casts a long shadow on the melting transition. This is clear from
(\ref{eiii}) and, by inference, from (\ref{eii}). In the
$\Phi$-ordered phase, there is an effective long range interaction
between (s) vortex lines which, in the $q\ll 1/\ell$ limit, takes the 
London-Biot-Savart form.\cite{blatterrmp}
To see this, note that,  in the $\Phi$-ordered ``Meissner" phase, our fictitious
``photon" is ``massive", i.e.,  the second (gradient) term in Eq. (\ref{eiii})
becomes:
\begin{equation}
\gamma_{\perp}\vert \langle\Phi\rangle \vert^2 {\bf S}_{\perp}^2 +
\gamma_{\parallel}\vert \langle\Phi\rangle \vert^2 {\bf S}_{\parallel}^2 ~~~,
\label{exxx}
\end{equation}
where $\langle\Phi\rangle$ is the order parameter associated
with the $\Phi$-order. As seen in Sec. V, we expect that, as long
as $\langle\Phi\rangle$ is finite, we can write some effective description
of such a state in terms of a ``line liquid" of field-induced vortices,
in the sense of Nelson.\cite{nelson} Long distance physics 
can be described through fluctuations in the density and ``currents"
of vortex lines, $\delta\rho ({\bf r})$ and ${\vec j}({\bf r}) = (j_x,j_y)$,
respectively.\cite{nelson}
The connection with our fictitious ``gauge" potential ${\bf S} ({\bf r})$ (\ref{eiii})
is (Appendix A):
\begin{equation}
(\nabla\times {\bf S})_{\parallel} \to 2\pi\delta\rho~~,
~~~(\nabla\times {\bf S})_{\perp} \to 2\pi {\vec j}~~~.
\label{exxxi}
\end{equation}
If we now reexpress (\ref{exxx})
in terms of $\delta\rho$ and ${\vec j}$ we get precisely the long distance part of the
London-Biot-Savart interaction between field-induced vortex lines:
\begin{equation}
4\pi ^2\gamma\vert \langle\Phi\rangle \vert^2 
\sum_{\bf q}\frac{\delta\rho ({\bf q})\delta\rho (-{\bf q}) +
{\vec j}({\bf q})\cdot\vec j(-{\bf q})}{q_{\perp}^2 + q_{\parallel}^2}~~,
\label{exxxii}
\end{equation}
where the continuity condition, 
${\vec q}\cdot {\vec j}({\bf q})=-q_{\parallel}\delta\rho ({\bf q})$,
is assumed. The anisotropy, suppressed in the above expression for
simplicity, can be straightforwardly restored.\cite{sardella}
This expression (\ref{exxxii}) for the effective interaction at long
distances is just what is obtained in the mean-field based approach,\cite{dodgson} but
with one {\em crucial difference}. The overall strength of the interaction
is not given by the mean-field {\em amplitude} squared of the superconducting
order parameter, but by $\vert\langle\Phi\rangle\vert^2\propto n_{\Phi}$, 
where $n_{\Phi}$ is 
the $\Phi$-superfluid density, whose physical meaning is apparent from 
reformulation (\ref{eii}) and gauge theory (\ref{eiii}). 

The immediate consequence of (\ref{exxxii}) 
is that the melting line, $T_m(H)$,
goes into $T_{c0}$, the true zero-field superconducting transition, as
$H\to 0$. This result is strongly suggested by all available numerical
simulations on the 3D XY model and arises naturally in the gauge theory
scenario. For all its apparent simplicity, this result
is not trivial: the mean-field based
theories of melting including only field-induced London vortices\cite{dodgson}
naturally lead to $T_m(H)\to T_c$, the {\em mean-field} transition
temperature, as $H\to 0$.\cite{footx,footxi} Furthermore,
the exponent $\nu$ has its
mean-field value $\nu_{\rm mf}=1/2$  and is {\em not} equal to $\nu_{xy}\sim 2/3$.
Therefore, the thermodynamics of the melting transition resulting from
such theories cannot satisfy the 3D XY scaling properties of Sec. IV.
This is a direct consequence of ignoring those very degrees of freedom
(vortex loops) which are primarily responsible for moving
the true {\em zero} field superconducting transition temperature from $T_c$ to
$T_{c0}$ and changing $\nu_{\rm mf}$ to $\nu_{xy}$ in the first place. 
This is a serious flaw and must be rectified in 
a proper theory of vortex lattice melting in the critical region.

As a first step, we attempt to remedy the situation by simply replacing, 
{\em by hand}, the mean-field amplitude squared with the true superfluid
density at {\em zero} field. This amounts to installing
$\vert\langle\Phi \rangle_{H=0}\vert^2$ instead of 
$\vert\langle\Phi \rangle\vert^2$ in Eq. (\ref{exxxii}). 
With the interaction fixed in this
fashion, we can then proceed to analyze the {\em same},
{\em finite} field model of Ref.\cite{dodgson}, still including only the field-induced
vortices. This is equivalent to having $T_{\Phi}(H)=T_{\Phi}(H=0)\equiv T_{c0}$,
as represented by the dashed vertical line in Fig. 1.
This procedure, arbitrary as it is within the framework of mean-field
based theories, seems to get us out of the above difficulties,
since now evidently $T_m(H)\to T_{c0}$ for $H\to 0$ and 
the exponent $\nu$ takes its true $H=0$
value, $\nu_{xy}\sim 2/3$.

However, things are not that simple: to understand why 
note that the above remedial procedure is in fact
{\em exact}, but only for a {\em single} field-induced line. For a {\em finite}
density of lines, 
the physical state of thermally-generated vortex loops and other
fluctuations described by $\Phi ({\bf r})$, which controls
the effective interaction between field-induced vortex
lines through $\vert\langle\Phi \rangle\vert^2$ in Eq. (\ref{exxxii}), 
is {\em itself} strongly affected by interactions with those
same field-induced lines. The effective coupling of these two interpenetrating
systems, vortex loops and lines in reformulation
(\ref{eii}), must be solved {\em self-consistently} at finite $H$: 
this is precisely what is accomplished in the gauge theory scenario (\ref{eiii})
for the long wavelength ($q\ll 1/\ell$) behavior. 
Clearly, a ``minimal" theory of vortex lattice melting in the critical region must involve
{\em both} the positional order parameter of the vortex lattice $\rho_G$ 
(or the original $\Psi$) {\em and} the new ``superconducting" order parameter
$\Phi$. The coupled equations governing the ($T, H$) dependence of these two order parameters
must be solved simultaneously and self-consistently near $T_m(H)$.

An important physical feature is expected to emerge from such a solution:
{\em the formation of vortex lattice is a phase transition involving simultaneous
ordering of both field-induced and thermally-generated degrees of freedom}.
This is illustrated with two qualitative points. First, at low fields
within the $\Phi$-ordered
state (Fig. 1), we can consider some effective ``line-liquid" description (\ref{exxxi}).
Right above $T_m(H)$, in the liquid phase, the self-consistent
solution gives $\langle\Phi\rangle =\langle\Phi\rangle_L$ to be inserted in the effective
interaction (\ref{exxxii}). Similarly, just below $T_m(H)$, we have
$\langle\Phi\rangle =\langle\Phi\rangle_S$. In general, however,
$\langle\Phi\rangle_L\not =\langle\Phi\rangle_S$! This result follows immediately
from the gauge theory scenario (\ref{eiii}) since positional correlations of s vortices
strongly influence $\langle\Phi\rangle$.  While both 
$\langle\Phi\rangle_L$ and $\langle\Phi\rangle_S$ are finite, as we cross $T_m(H)$,
there is a discontinuous change in the average density and size of vortex loop and
``overhang" excitations, resulting in different effective interaction (\ref{exxxii})
on two sides of the melting line (Fig. 4). The entropy jump at melting, $\Delta S$,
will receive significant contribution from such excitations.
In fact, while $\Delta S\to 0$ as $H\to 0$, the 
{\em critical fluctuations} greatly {\em enhance} $\Delta S$ over
the configurational entropy of {\em field-induced lines}.  
This provides natural explanation\cite{zt}
for the excess entropy at $T_m(H)$
observed in low field thermodynamics.\cite{zeldov,schilling}
It should be stressed that this effect is {\em different} from 
the ``microscopic entropy" contribution discussed by
Hu and MacDonald\cite{hu} (see also Ref.\cite{dodgson}). Such entropy arises from
the {\em electronic} degrees of freedom and 
is reflected in the $T$-dependence of our GL coefficients (\ref{ei}). This 
contribution is important in the high-field, LL regime.\cite{zta,hu}
In the 3D XY critical region,
however, such $T$-dependence is a {\em minor} effect since it involves
$T_c$ and not the true transition temperature $T_{c0}$. The entropy 
contribution discussed here, arising from $T$-dependence of $|\langle\Phi\rangle|^2$
in Eq. (\ref{exxxii}) and $\langle\Phi\rangle_L\not =\langle\Phi\rangle_S$, is
due to degrees of freedom of the {\em superconducting order parameter} itself
and must be part of any proper theory of melting.

Second, at higher fields, 
the self-consistent solution for $\rho_G$ (or $\Psi$) and $\Phi$ leads to
a rapid suppression of $T_{\Phi}(H)$ far below $T_{c0}$, and its subsequent
merging with $T_m(H)$ for $H>H_Z$ (Fig. 1). This exposes a large region of the phase
diagram where the Abrikosov vortex lattice melts directly into the true normal state.
In this case $\langle\Phi\rangle _L =0$, while 
$\langle\Phi\rangle _S$ might still be close to its mean-field
value.  Such dramatic difference in the nature of these two phases, with the
solid being rather unremarkably mean-field like, and the liquid right
above the melting line exhibiting very strong fluctuations
even at rather low fields, with
vortex lines winding {\em both} along the field {\em and} in the
perpendicular directions, 
is evident in new simulations by Nguyen and Sudb{\o}\cite{sudbo,sudbopc}
Note that such situation {\em never} arises in the mean-field based theories
with only field-induced vortices,\cite{dodgson} even after the application
of our remedial procedure, since we would still have 
$\vert\langle\Phi \rangle_{S}\vert^2=
\vert\langle\Phi \rangle_{L}\vert^2=
\vert\langle\Phi \rangle_{H=0}\vert^2$ in Eq. (\ref{exxxii}). 
As argued in Sec. V, it does not appear possible 
to write an effective description of the true normal state in terms
of field-induced degrees of freedom only. 
For example, we could start again with our remedial procedure and argue that, even though
it fails for $q\ll 1/\ell$, it still describes
the effective interaction of field-induced vortices for $q\sim 1/\ell$, which
is what matters most at $T_m(H)$. 
However, within such a ``line liquid" description,
I do not see any simple way by
which one could account for the part of $\Delta S$ associated
with a discontinuous change in windings 
across $T_m(H)$.
In this region of higher fields, where $T_{\Phi}(H)=T_m(H)$, our gauge theory
is becoming less and less ``type-II" and amplitude fluctuations are
becoming stronger in the GL theory. It is likely that the low-field
description discussed here and the LL theories now have an equal chance of providing
reliable theory of the melting transition: both routes, however, are
certain to be most challenging.

\section*{Acknowledgments}
The author would like to thank L. N. Bulaevskii, G. W. Crabtree,
M. Dodgson, M. Friesen, I. F. Herbut, R. Ikeda, 
K. Kadowaki, A. Koshelev, M. A. Moore, T. Nattermann, 
A. K. Nguyen, S. Pierson, S. Ryu, D. Stroud, U. C. Taeuber, S. Teitel, 
O. T. Valls, A. van Otterlo,
and particularly A. Sudb{\o} for discussions and encouragement.
This research was supported in part by the NSF
grant DMR-9415549. 


\end{multicols}

\section*{Appendix A: Derivation of the Gauge Theory}
I present here a derivation of the gauge theory (\ref{eiii}). An
abbreviated version is found in Ref.\cite{zt}. For simplicity,
I consider the isotropic case, $\gamma_{\parallel}=\gamma_{\perp}=\gamma$.

Within the ``helium model",\cite{popov} the partition function of the superconductor
(\ref{ei}) can be written as:
$$Z=\sum\cdots\sum_{N^{({\bf\omega})}}\prod_{\bf\omega}\frac{1}{N^{({\bf\omega})}!}
\prod_{l_{\bf\omega}=1}^{N^{({\bf\omega})}}\int {\cal D}
{\bf x}_{l_{\bf \omega}}[s_{l_{\bf\omega}}]\exp\bigl(
-\frac{F_v}{T}\bigr);~ F_v = \gamma\langle |\Psi |^2\rangle 
\int_{\cal C} d^3r
|\nabla\varphi ({\bf r}) - \frac{2e}{c}{\bf A}|^2
+ \sum_{\bf\omega}\sum_{l_{\bf\omega}=1}^{N^{({\bf\omega})}}
\int ds_{l_{\bf\omega}}E^{(1)}(\{{\bf x}_{l_{\bf \omega}}[s_{l_{\bf\omega}}]\})$$
\begin{equation}
+ \frac{1}{2}\sum_{\bf\omega}\sum_{\bf\omega'}
\sum_{l_{\bf\omega}=1}^{N^{({\bf\omega})}}
\sum_{l_{\bf\omega'}=1}^{N^{({\bf\omega'})}}
\int ds_{l_{\bf\omega}}
\int ds_{l_{\bf\omega'}}\bigl\{
V^{(2)}_0(|{\bf x}_{l_{\bf \omega}}[s_{l_{\bf\omega}}]-
{\bf x}_{l_{\bf \omega'}}[s_{l_{\bf\omega'}}]|) +
\frac{d{\bf x}_{l_{\bf \omega}}}{ds_{l_{\bf\omega}}}\cdot
\frac{d{\bf x}_{l_{\bf \omega}}}{ds_{l_{\bf\omega'}}}
V^{(2)}_1(|{\bf x}_{l_{\bf \omega}}[s_{l_{\bf\omega}}]-
{\bf x}_{l_{\bf \omega'}}[s_{l_{\bf\omega'}}]|)\bigr\} ~~~,
\label{ebi}
\end{equation}
where
\begin{equation}
\nabla\times {\bf A}={\bf H}~~;~~
\nabla\cdot\nabla\varphi ({\bf r})=0~~;~~\nabla\times\nabla\varphi ({\bf r})
=2\pi \sum_{\bf\omega}\sum_{l_{\bf\omega}=1}^{N^{({\bf\omega})}}\int _{\cal L}
d{\bf x}_{l_{\bf\omega}}
\delta ({\bf r} - {\bf x}_{l_{\bf\omega}}[s_{l_{\bf\omega}}])~~~.
\label{ebii}
\end{equation}

The partition function (\ref{ebi}) is a 3D counterpart of the familiar
representation of the (continuum) 2D XY model in terms of its point-like
topological excitations, vortices and anti-vortices. In 3D, the relevant excitations
are loops/lines of vortices, classified by their global topology. Only
vortex paths of unit vortex ``charge" are considered since they are 
the important excitations in the critical region.
The lattice regularization of (\ref{ebi}) is a gas of non-intersecting oriented paths
on a lattice which are either closed or can start/end only on sample surfaces.
The lattice spacing is set by the characteristic ``bending length"
of vortex lines.  Each individual step along a 
path takes a given energy to
create and can be either up or down along $x,y$ or $z$ axis. These
steps represent vortex segments and have a long range ``directional" Coulomb
interaction, operating only between the steps going along the same axis.
This is the lattice version of the ``Biot-Savart" interaction between vortex loops/lines
of the continuum model. The background free energy, composed of the
uniform condensation energy and ``spin-wave" contributions, is not included
explicitly.

The summation in (\ref{ebi}) runs over all distinct 
configurations of vortex line excitations
of arbitrary length and shape. The index ${\bf\omega}$ denotes 
different classes of oriented loops/lines
which are distinguished by their global topology. For example, for periodic
boundary conditions, when only closed loops are present,
${\bf\omega}=(m^{\pm}_x;m^{\pm}_y;m^{\pm}_z)$, where
$m^{\pm}_{x,y,z}=0,\pm 1,\pm 2,\dots$ denote the winding numbers of 
a given loop around $x,y,z$ directions. {\em Finite} closed
loops correspond to $m^{\pm}_x=m^{\pm}_y=m^{\pm}_z=0$. Similarly, for free (periodic)
boundary conditions along $z$ ($x,y$) direction,
${\bf\omega}=(m_x;m_y;0,0)$ denotes loops that wind in the $x$ ($y$) direction
while ${\bf\omega}=(0,0;0,0;0,0)$ again denotes finite closed loops.
In addition, there are vortex (antivortex) paths that traverse the system
from $z=0$ to $z=L_z$ and ``half-loops" which originate and terminate
at the {\em same} $z=0$ or $z=L_z$ surface. 
$\int {\cal D}{\bf x}_{l_{\bf \omega}}[s_{l_{\bf\omega}}]$ 
represents summation over all configurations of a 
given loop/line $l_{\bf\omega}$ consistent with its global topology.
${\cal C}\{ {\bf x}_{l_{\bf \omega}}[s_{l_{\bf\omega}}]\}$ in the
first term of $F_v$ signifies that
the integral of the gradient energy over the system excludes well-defined core
regions associated with a given configuration of loops/lines. 
The second and third term
represent core contributions: $E^{(1)}=E_c + 
E_b(\{ {\bf x}_{l_{\bf \omega}}[s_{l_{\bf\omega}}]\} )$
is a ``single-particle"
term, with $E_c$ and $E_b$ corresponding to the core line and bending energies,
respectively.  $V_{0,1}^{(2)}(|{\bf r}-{\bf r'}|)$
denotes ``two-particle" effects of core overlap. These ``two-particle"
terms describe both the energy cost of core overlap and 
the entropic effects of keeping vortex and anti-vortex segments 
from annihilating each other.
``Multi-particle" terms, arising from simultaneous overlap of more
than two cores, can be neglected in the extreme type-II regime, where 
the average core size $a$ is small compared to the
average separation between vortex segments.  
Core contributions, like $E_c$, $E_b$ and 
$V_{0,1}^{(2)}(|{\bf r}-{\bf r'}|)$, can be computed in a specific microscopic
model of vortex lines.\cite{pit,chaikin} Their precise form is not needed for 
our present purposes since we will return to the GL
representation at the end of this Appendix;
it suffices to know that $E_c$ and $E_b$ are finite 
and the ``interaction" $V_{0,1}^{(2)}(|{\bf r}-{\bf r'}|)$
is short-ranged, of order $a$, and repulsive on average.
Without loss of generality, we could set
$V_{0,1}^{(2)}(|{\bf r}-{\bf r'}|)\to
V_{0,1}\delta ({\bf r}-{\bf r'})$.

With the uniform magnetic field present, the overall vortex ``charge" neutrality
demands that every configuration contains $N_{\Phi}$
field-induced vortex paths going from $z=0$ to $z=L_z$. This fact is used to
observe that the low temperature expansion of $Z'$ (\ref{eii}) in terms
of topological defects of the new order parameter $\Phi ({\bf r})$ and
s vortices, is the same as (\ref{ebi}) except for different prefactors:
$1/(N_{\Phi}+N_a)!$ in $Z$ versus $1/N_{\Phi}!N_a!$ in $Z'$, where
$N_a$ is the number of thermally-generated infinite vortex-antivortex
paths which extend from $z=0$ to $z=L_z$. This leads to a difference
in entropy between two representations $\Delta S\sim T\ln [(N_{\Phi}+N_a)!/N_{\Phi}!N_a!]$.
$\Delta S$ scales as $L_xL_y$, in contrast to the full entropy
which goes as $L_xL_yL_z$. Consequently, in the thermodynamic limit,
$\Delta S$ does not affect the free energy per unit volume. In particular,
within the $\Phi$-ordered state, $Z$ (\ref{ei}) and $Z'$ (\ref{eii})
have identical expansions:
$$Z=\sum_{N^{(0)}=0}^{\infty}\frac{1}{N^{(0)}!}\frac{1}{N_{\Phi}!}
\prod_{l_0=1}^{N^{(0)}} \oint {\cal D}_b{\bf x}_{l_0}[s_{l_0}]
\prod_{i=1}^{N_{\Phi}}
\int {\cal D}_b {\bf r}_i[s_i]\exp\bigl(
-\frac{F_v}{T}\bigr)~~~;$$
$$ F_v = \gamma\langle |\Psi |^2\rangle 
\int_{\cal C} d^3r
|\nabla\phi ({\bf r}) + {\bf U} - \frac{2e}{c}{\bf A}|^2
+ \sum_{l_0=1}^{N^{(0)}} E_c\int ds_{l_0}
+ \sum_{i=1}^{N_{\Phi}} E_c\int ds_i+$$
\begin{equation}
\frac{1}{2}\int d^3r\int d^3r'\bigl\{
[d_0({\bf r}) + d_s({\bf r})]
[d_0({\bf r'}) + d_s({\bf r'})]
V^{(2)}_0(|{\bf r}-{\bf r'}|) +
[{\bf n}_0({\bf r}) + {\bf n}_s({\bf r})]\cdot
[{\bf n}_0({\bf r'}) + {\bf n}_s({\bf r'})]
V^{(2)}_1(|{\bf r}-{\bf r'}|)\bigr\}~.
\label{ebiii}
\end{equation}
Here $\{{\bf x}_{l_0}[s_{l_0}]\}$ denote {\em finite} closed loops of arbitrary
length and shape. Half-loops attached to $z=0$ and $z=L_z$ surfaces are 
not included since they do not matter for the bulk thermodynamics.
The symbol ${\cal D}_b$ indicates that
the ``single-particle" bending energy
$(E_b(\{ {\bf x}_{l_{\bf \omega}}[s_{l_{\bf\omega}}]\} ))$
has been absorbed into the measure of the path integral. 
``Densities" and ``currents" $d_{0,s}$ and ${\bf n}_{0,s}$ are defined as:
\begin{equation}
\{ d_0,{\bf n}_0\}({\bf r})=\sum_{l_0=1}^{N^{(0)}}\int ds_{l_0}
\{1,\frac{d{\bf x}_{l_0}}{ds_{l_0}}\}
\delta ({\bf r} - {\bf x}_{l_0}[s_{l_0}])~~~;~~~
\{d_s,{\bf n}_s\}({\bf r})=\sum_{i=1}^{N_{\Phi}}\int ds_i
\{1,\frac{d{\bf r}_i}{ds_i}\} \delta ({\bf r} -{\bf r}_i[s_i])~~~,
\label{ebiv}
\end{equation}
and
\begin{equation}
\nabla\cdot\nabla\phi ({\bf r})=0~~,~~\nabla\times\nabla\phi ({\bf r})
=2\pi {\bf n}_0({\bf r})~~;~~
\nabla\cdot {\bf U}({\bf r})=0~~,~~\nabla\times {\bf U} ({\bf r})
=2\pi {\bf n}_s({\bf r})~~.
\label{ebv}
\end{equation}

To understand the problem at low fields, we first consider the $H=0$ situation.
The superconducting (Meissner) state is described by the ${\bf H}\to 0$,
$N_{\Phi}\to 0$ limit of (\ref{ebiii}):
$$Z(H=0)=\sum_{N^{(0)}=0}^{\infty}\frac{1}{N^{(0)}!}
\prod_{l_0=1}^{N^{(0)}} \oint {\cal D}_b{\bf x}_{l_0}[s_{l_0}]
\bigl( -\frac{F_v(0)}{T}\bigr)~~~;~~~
F_v(0) = \gamma\langle |\Psi |^2\rangle
\int_{\cal C} d^3r
|\nabla\phi ({\bf r})|^2
+ \sum_{l_0=1}^{N^{(0)}} E_c\int ds_{l_0}+$$
\begin{equation}
+\frac{1}{2}\int d^3r\int d^3r'\bigl\{
d_0({\bf r})d_0({\bf r'})
V^{(2)}_0(|{\bf r}-{\bf r'}|) +
{\bf n}_0({\bf r})\cdot {\bf n}_0({\bf r'})
V^{(2)}_1(|{\bf r}-{\bf r'}|)\bigr\}~~~.
\label{ebvi}
\end{equation}
The gradient term in $F_v(0)$ can be decoupled by a {\em dual}
gauge field ${\bf A}_d({\bf r})$, in the Coulomb gauge $\nabla\cdot {\bf A}_d=0$:
\begin{equation}
\exp\bigl[-\frac{\gamma\langle |\Psi |^2\rangle}{T}
\int_{\cal C} d^3r
|\nabla\phi ({\bf r})|^2\bigr]\to
\int {\cal D}{\bf A}_d({\bf r})
\exp\bigl[\int d^3r\bigl(-i{\bf n}_0\cdot {\bf A}_d -
\frac{1}{2e^2_d}(\nabla\times{\bf A}_d)^2\bigl)\bigr]~~,
\label{ebvii}
\end{equation}
where $e^2_d= 8\pi ^2\gamma\langle |\Psi |^2\rangle /T$ is a dual charge.
Eqs. (\ref{ebvii}) and (\ref{ebvi}) have an appearance of a path integral
over trajectories of {\em relativistic} charged quantum particles in
a $2+1$ dimensional Euclidean space ($z$ being the imaginary time) coupled to
the ``electromagnetic" gauge potential ${\bf A}_d$ and interacting via
short-range ``density-density" and ``current-current" interactions. 
$Z(H=0)$ describes the {\em vacuum} structure of such 
``electrodynamics", with our vortex loops corresponding to worldlines
of relativistic quantum bosons and describing {\em virtual} particle-antiparticle
creation and annihilation processes in the vacuum. This similarity can
be exploited further by using the particle-field duality to define
the field-theory version\cite{parisi} of $Z(H=0)$ ({\ref{ebvii},\ref{ebvi}):
\begin{equation}
\int {\cal D}\Psi_d ({\bf r})\int {\cal D}{\bf A}_d({\bf r})
\exp\bigl\{-\int d^3r\bigl[m_{\Psi}^2|\Psi_d |^2 + |(\nabla -i{\bf A}_d)\Psi _d|^2
+\frac{1}{2}g_0|\Psi_d|^4+ \frac{1}{2q^2_d}(\nabla\times{\bf A}_d)^2
+\frac{M_d^2}{2}{\bf A}_d^2\bigl]\bigr\}~~.
\label{ebviii}
\end{equation}
$\Psi_d ({\bf r})$ is a field operator of these relativistic bosons.
The gradient term in the action has been rescaled into dimensionless form so that 
short-range repulsion and dual charge assume their canonical dimensions:
$V_0 \delta ({\bf r}-{\bf r'})\to g_0 \delta ({\bf r}-{\bf r'})$,
$[g_0]=$ (length)$^{-1}$; 
$e_d^2\to q_d^2$, $[q_d^2]=$ (length)$^{-1}$. $V_1$ has been dropped since it
is irrelevant in the long wavelength limit. Its effect on critical behavior 
can be incorporated into the bare values of $g_0$, $q_d^2$, and $m_{\Psi}^2$.
Finally, the ``bare mass" of ${\bf A}_d$, $M_d$, is {\em absent} in our problem ($M_d=0$).
Finite $M_d$ reflects the presence of a gauge field minimally coupled
to the original, superconducting order parameter. For example, if the condition
$\kappa\to\infty$ is relaxed and the real electromagnetic screening is restored in (\ref{ei}), 
$M_d^2\to \mu_0 e^2/\pi$, where $e$ and $\mu_0$ are the real charge and
magnetic permeability, respectively.

The expression (\ref{ebviii}) forms the basis for the ``dual"
picture of the 3D XY critical behavior.\cite{dasgupta,kleinert} 
In this picture, we are
viewing vortices as primary objects and their field-operator, $\Psi_d({\bf r})$, as
our order parameter, instead of the original $\Psi ({\bf r})$. 
We can think of $\Psi ({\bf r})$ in Eq. (\ref{ei}) as being the 
field-operator describing creation
and annihilation of Cooper pairs. In the GL theory, with $H=0$  (\ref{ei}),
Cooper pairs have only short range interactions and it suffices to keep
only the quartic term, describing the point-like repulsion, since the
rest is irrelevant for the critical behavior. In contrast,
the {\em vortex} excitations of  $\Psi$ interact via long range London-Biot-Savart forces,
mediated by massless ($M_d=0$) ${\bf A}_d$ (\ref{ebviii}).
Next, we can convert our neutral GL theory (\ref{ei}) 
into the one with finite real charge, $e$,
by introducing the fluctuating vector potential, ${\bf A}$, 
as well as the electromagnetic field energy,
$(1/8\pi\mu_0 e^2)(\nabla\times {\bf A})^2$. Now, 
it is the Cooper pairs that have long range interactions, mediated by ${\bf A}$, but
the vortices in $\Psi$ interact only through short range forces, due
to the electromagnetic screening inducing a finite $M_d=\mu_0 e^2/\pi$ in (\ref{ebviii}). 
Precisely in 3D, there is an exact duality, at least on a 
lattice,\cite{dasgupta} between the
form of this interaction for vortices and for Cooper pairs.\cite{kleinert} 
This is what ``inverted" stands for in the ``inverted 3D XY" model. In the dual
language of $\Psi_d$, it is the low temperature Meissner phase of the original
superconductor that is symmetric ($\langle\Psi_d\rangle =0$), 
while the high temperature normal metal 
is the ``broken symmetry" state of the dual theory 
($\langle\Psi_d\rangle\not =0$). So, there is an
inversion of the temperature axis. However, it is still the same symmetry, $U(1)$,
that is being broken and this implies that the {\em thermodynamic} exponent 
should be the same, $\nu = \nu_{xy}$, {\em both} for $e = 0$ {\em and}
for $e\not =0$. It is important to stress that the ``inverted 3D XY"
behavior of a charged-superfluid remains  {\em different} from
the 3D XY critical behavior of a neutral-superfluid, described by (\ref{ei})
with $H=0$, since they are associated with two different critical points.
For example, the anomalous dimension exponent, $\eta _{\Psi}$, of $\langle\Psi (0)
\Psi ^*({\bf r})\rangle$, will be {\em different} in
two cases, $e=0$ and $e\not =0$.

Eq. (\ref{ebviii}) describes the ``true vacuum" 
state $|0\rangle$ of a Euclidean relativistic field
theory. Particle (antiparticle) excitations are massive, 
$m_{\Psi}\sim 1/\xi_{d}\sim 1/\Lambda_{\Psi}$,
where $\xi_{d}$ is the dual correlation length associated with $\Psi_d$
and $\Lambda_{\Psi}$ measures the average loop size. 
The actual value of $m_{\Psi}$ reflects both the cost in energy
and gain in entropy arising from large thermally-generated vortex loops
in the original problem. As we approach $T_{c0}$ from below, $m_{\Psi}\to 0$,
and we enter the ``false vacuum" state $|f\rangle$ of the theory (\ref{ebviii}).
Particle (antiparticle) excitations are now massless and infinite vortex
paths proliferate across the system, as depicted in Fig. 2.
This ``false vacuum" is just the normal metallic state. If we
introduce the particle (antiparticle) number operators, $\hat N_{P,A}$,
$|0\rangle$ is an eigenstate of $\hat N_{P,A}$,
$\hat N_P|0\rangle = 0|0\rangle$ and  $\hat N_A|0\rangle = 0|0\rangle$.
In contrast, $|f\rangle$ is {\em not} an
eigenstate of $\hat N_{P,A}$, and contains a {\em finite} average  number of
(anti)particles, $\langle f|\hat N_{P,A}|f\rangle = N_{P,A}\not =0$.
Both $|0\rangle$ and $|f\rangle$, however, are eigenstates
of the total vorticity operator $\hat N_P -\hat N_A$ with eigenvalue 0,
which insures $N_P=N_A$. For $H$ small but finite, the ground state of (\ref{ebviii}) must still
be an eigenstate of  $\hat N_P -\hat N_A$ but now the eigenvalue is
$N_{\Phi}$. Starting from $|0\rangle$, such ground state $|\Phi\rangle$ is naturally 
constructed by introducing $N_{\Phi}$ {\em massive} particles into the
true, stable vacuum. We then have $\hat N_P|\Phi\rangle =N_{\Phi}|\Phi\rangle$ and
$\hat N_A|\Phi\rangle =0|\Phi\rangle$. On the other hand, starting
from the ``false vacuum" $|f\rangle$, the ground state $|n\rangle$
is formed by having additional $N_{\Phi}$ particles added to
an already present finite average number of {\em massless} particle-antiparticle
pairs. $|n\rangle$ is not an eigenstate of $\hat N_{P,A}$
and satisfies $\langle n|\hat N_{P,A}|n\rangle = N_{P,A} \not =0$,
where now $N_P-N_A=N_{\Phi}$. At $T_{\Phi}(H)$, a phase transition
takes place between these two different types of ground state,
$|\Phi\rangle$ and $|n\rangle$, driven by the change in $U(1)$ symmetry
of the vortex system.

We now return to Eq. (\ref{ebiii}). A finite density of s vortices 
produces two main effects on the loop ``expansion" as we approach
$T_{\Phi}(H)$ from below. First, there is a long-range
interaction between loops and s vortices which will influence the long-range
correlations among the loops themselves. Second, there is a short-range effect
of s vortices suppressing certain configurations of loops, 
through mutual contact interactions (an ``excluded volume" effect). 
Intuitively, one expects the long-range effect to be essential for the critical
behavior at low fields. Furthermore, the short range effect should be weak
since the total number of vortex segments connected to s vortices forms a
minority of all vortex segments. Based on these observations, we devise
the following strategy: according to our basic assumption ii) and results
of Sec. V, the system of s vortices below $T_{\Phi}(H)$ can be 
viewed as equivalent, in the long wavelength limit,
to an effective system of {\em non-relativistic} 2D quantum bosons in its
{\em superfluid} state, in the sense of Nelson.\cite{nelson} 
The collective modes of such a system are its ``density" and ``current"
fluctuations, which is precisely how the loops couple to s vortices.
The long wavelength effective action of these collective modes is
described by two coupling constants, $m_s/n_s$ and $c_s^2$, where, in
this boson analogy, $n_s$, $m_s$ and $c_s$ are the superfluid density, mass and speed of
``sound", respectively.  We could compute these quantities explicitely, 
by starting sufficiently below $T_{\Phi}(H)$. Here, however, we 
treat them as general parameters
which characterize the long wavelength fluctuations of s vortices
and whose ultimate values can be determined through their direct
connection with the components of the helicity modulus tensor (Sec. V and Appendix B).
We assume that the effect of mutual contact interaction between
loop and s vortex subsystems, measured by some interaction strength
$V_{\rho}$, can be fully included into these ultimate values of 
$m_s/n_s$ and $c_s^2$, as well as into a renormalized loop-loop contact
interaction, $\tilde V_{0}$ and the loop core line energy $\tilde E_c$. 
This amounts to reexpressing the short-range interaction in (\ref{ebiii}) as:
\begin{equation}
\frac{1}{2}\int d^3r\bigl\{
\tilde V_0d_0^2({\bf r}) + 
2V_{\rho}\delta d_0({\bf r})\delta d_s({\bf r})
+ V_{\rho\rho}d_s^2({\bf r})\bigr\}~~;
~~\tilde E_c = E_c + V_0\langle d_s ({\bf r})\rangle~~,
~~E_s = E_c + V_0\langle d_0 ({\bf r})\rangle~~,
\label{ebix}
\end{equation}
while the core line energies of loops and s vortices
are also renormalized to include the average ``excluded volume" effect.
$E_s$ and $V_{\rho\rho}$ will shortly be subsumed into $m_s/n_s$ and $c_s^2$.
We set $V_{\rho}\to 0$ in (\ref{ebix}) and proceed to derive the long distance
description of (\ref{ebiii}). Indeed, we find that the long-range interactions
lead to a major change in the behavior of the system once ${\bf H}$ is finite:
most importantly, the long-range interactions
between vortex loops are ``screened" by fluctuations of s vortex lines. The amount
of ``screening" is determined by $m_s/n_s$ and $c_s^2$.
Once the critical behavior associated with this ``screening" has been understood,
we reinstate the residual contact interaction between loop and
s vortex subsystems and test for consistency. 
We find, both within the $\epsilon$ expansion and perturbative RG in fixed
dimension $D=3$, that such residual coupling is irrelevant for $3\leq D<4$, i.e.,
it does not lead to any new relevant terms in the effective action, apart from
those already present. Our procedure is therefore self-consistent.

The first step is to decouple the gradient term in (\ref{ebiii}) by using ${\bf A}_d$,
except that now ${\bf n}_0\to {\bf n}_0 + \Delta {\bf n}_s$
in Eq. (\ref{ebvii}) ($\Delta {\bf n}_s$ is defined above Eq. (\ref{eiii})).
The s vortex part of (\ref{ebiii}) becomes:
\begin{equation}
\frac{1}{N_{\Phi}!}
\prod_{i=1}^{N_{\Phi}}
\int {\cal D}_b {\bf r}_i[s_i]
\exp\bigl\{
- \sum_{i=1}^{N_{\Phi}} \frac{E_s}{T}\int ds_i
- \frac{V_{\rho\rho}}{2T}\int d^3r d_s^2({\bf r}) -
i\int d^3r \Delta {\bf n}_s({\bf r})\cdot {\bf A}_d({\bf r}) \bigr\}~~.
\label{ebx}
\end{equation}
This part is now reexpressed in terms of s vortex ``density" and ``current" fluctuations.
Sufficiently below $T_{\Phi}(H)$, the overhangs are small and s vortex lines
are almost directed. We can use a non-relativistic boson analogy of Nelson\cite{nelson}
and replace $\int ds_i \approx \int_0^{L_z}dz\bigl(1+ \frac{1}{2}\bigl(\frac{d\vec r_i}
{dz}\bigr)^2\bigr)$, where $\vec r_i (z)=(x_i(z),y_i(z))$, while absorbing the
effect of small overhangs into the effective mass, $m_s\sim E_s/T$.
Similarly, $d_s({\bf r})\to {\bf n}_{s\parallel}({\bf r})\to\sum_i\delta (\vec r - \vec r_i(z))$
and ${\bf n}_{s\perp}({\bf r})\to\sum_i(d\vec r_i/dz)\delta (\vec r - \vec r_i(z))$.
We then introduce a field operator of these non-relativistic bosons, $\Psi_s (\vec r,z)$,
replace the path integral in (\ref{ebx}) with the functional integral
over $\Psi_s$ (periodic or free boundary conditions give the same result in the
thermodynamic limit), and follow the standard procedure\cite{popov} for deriving the
``hydrodynamic" action of s vortices:\cite{zt} 
the phase $\varphi_s (\vec r,z)$ and amplitude $\pi_s (\vec r,z)$ are
introduced via $\Psi_s=|\Psi_s|\exp(i\varphi_s)$ and
$\pi_s=|\Psi_s|^2-\langle|\Psi_s|^2\rangle$, where $\langle|\Psi_s|^2\rangle$
is evaluated self-consistently. After integration over $\pi_s$,
the ``hydrodynamic" action for the phase becomes:
\begin{equation}
\int d^3r\bigl\{
\frac{n_s}{2m_sc_s^2}(\nabla_{\parallel}\varphi_s -{\bf A}_{d\parallel})^2
+ \frac{n_s}{2m_s}(\nabla_{\perp}\varphi_s -{\bf A}_{d\perp})^2
+ (\cdots )\bigr\}~~~,
\label{ebxi}
\end{equation}
where $(\cdots )$ denotes higher powers of $\nabla\varphi_s -{\bf A}_d$
and higher order derivatives. The ``mean-field" part of the action is
absorbed into the background. The action (\ref{ebxi}) is
decoupled using real fields $\delta\rho ({\bf r})$
and $\vec j ({\bf r})= (j_x,j_y)$:
\begin{equation}
\to \int {\cal D}\delta\rho ({\bf r})
\int {\cal D}\vec j({\bf r})\exp\bigl\{
\int d^3r\bigl[
-i\delta\rho {\bf A}_{d\parallel} - i{\vec j}\cdot{\bf A}_{d\perp}
+i\delta\rho \nabla_{\parallel}\varphi_s + i{\vec j}\cdot\nabla_{\perp}\varphi_s
-\frac{m_sc_s^2}{2n_s}\delta\rho^2
-\frac{m_s}{2n_s}{\vec j}^2\bigr]\bigl\}~~~,
\label{ebxii}
\end{equation}
which finally results in Eq. (\ref{ebiii}) 
expressed in terms of the true collective modes of the superfluid s vortex system, 
its ``density" $\delta\rho$ and ``current" $\vec j$ fluctuations:
$$Z\to\sum_{N^{(0)}=0}^{\infty}\frac{1}{N^{(0)}!}
\prod_{l_0=1}^{N^{(0)}} \oint {\cal D}_b{\bf x}_{l_0}[s_{l_0}]
\int {\cal D}\delta\rho ({\bf r})
\int {\cal D}\vec j({\bf r})
\int {\cal D}{\bf A}_d({\bf r})
\exp (-S)~~~;~~~ S = \sum_{l_0=1}^{N^{(0)}} \tilde \frac{E_c}{T}\int ds_{l_0}+$$
\begin{equation}
+\int d^3r\bigl\{
\frac{\tilde V_0}{2T}
d_0^2({\bf r}) +
i{\bf n}_0\cdot {\bf A}_d +
i\delta\rho {\bf A}_{d\parallel} + i{\vec j}\cdot{\bf A}_{d\perp} 
+\frac{m_sc_s^2}{2n_s}\delta\rho^2
+\frac{m_s}{2n_s}{\vec j}^2
+{\cal W}(\delta\rho,\vec j)
+\frac{1}{2e^2_d}(\nabla\times{\bf A}_d)^2\bigr\}~~,
\label{ebxiii}
\end{equation}
where the functional integral must be appended by the continuity condition,
$\nabla_{\parallel}\delta\rho +\nabla_{\perp}\cdot\vec j =0$, which
follows from the integration over $\varphi_s$ in (\ref{ebxii}).
${\cal W}$ arises from $(\cdots )$ terms in (\ref{ebxi}) and
contains powers higher than quadratic
in $\delta\rho$ and $\vec j$, as well as assorted derivatives. In
particular, {\em odd} powers of $\delta\rho$ are present, like
$\sim \delta\rho^3$, reflecting the $\Delta {\bf n}_{s\parallel} 
\to -\Delta {\bf n}_{s\parallel}$ asymmetry
of the original problem (\ref{ebiii}). Such higher order asymmetric terms are essential for
description of the melting transition but, as shown below,
are irrelevant at the $\Phi$-transition, provided the latter is continuous.
Note that the above relatively simple dependence
of (\ref{ebxiii}) on $\delta\rho$ and $\vec j$ holds only at distances
$>\ell$. This is precisely what we are interested in as we approach $T_{\Phi}(H)$.
Eq. (\ref{ebxiii}) captures an essential effect of a finite field on the
loop ``expansion": fluctuations of field-induced s vortex lines result
in the ``screening" of the ``Biot-Savart" interaction between
the loops. This ``screening" is manifested by ${\bf A}_d$ gaining a finite ``mass",
$M^2_{\parallel}\sim n_s/m_sc_s^2$, 
$M^2_{\perp}\sim n_s/m_s$, after integration over 
$\delta\rho$ and $\vec j$. The effect of finite magnetic field in the
original problem (\ref{eii}) is now stored in the finite values of
$M^2_{\parallel}$ and $M^2_{\perp}$.  As one attempts to create 
a small density of very large loops in (\ref{ebxiii}),
upon approaching $T_{\Phi}(H)$ from below,
their effective line tension and mutual interactions will be essentially
influenced by such ``screening".

The importance of this ``screening" mechanism is particularly apparent in the
dual representation (\ref{ebviii}). The action (\ref{ebxiii}) becomes:
\begin{equation}
\int d^3r\bigl[m_{\Phi}^2|\Phi_d |^2 + |(\nabla -i{\bf A}_d)\Phi _d|^2
+\frac{\tilde g_0}{2}|\Phi_d|^4+ 
i\delta\rho A_{dz} + i{\vec j}\cdot\vec A_d +\frac{m_sc_s^2}{2n_s}\delta\rho^2
+\frac{m_s}{2n_s}{\vec j}^2
+{\cal W}
+ \frac{1}{2q^2_d}(\nabla\times{\bf A}_d)^2\bigl]\bigr\},
\label{ebxiv}
\end{equation}
where the meaning of $\Phi_d ({\bf r})$, 
$m_{\Phi}$, and $\tilde g_0$ is evident
in light of the discussion surrounding Eq. (\ref{ebviii}).
For simplicity, I suppress the anisotropy in the
second (gradient) term of (\ref{ebxiv}) which generically
arises (even if $\gamma_{\parallel}=\gamma_{\perp}$ in (\ref{ei}))
from the interaction of loops with s vortices.
The finite mass of ${\bf A}_d$, generated by 
the integration over $\delta\rho$ and $\vec j$, 
``cuts off" the long-range ``Biot-Savart" 
interactions present in zero field (\ref{ebviii}).  
This ``screening" causes decoupling of the dual gauge field
and transformes the critical behavior from 
a ``charged" (\ref{ebviii}) to a ``neutral" 
(\ref{ebxiv}) dual superfluid. This is the inverted (anisotropic) 3D XY
behavior\cite{dasgupta,kleinert} of the dual theory, 
hinting at the presence of a {\em massless} gauge field in the 
{\em original} ``superconducting" formulation (\ref{eiii}), as
discussed below Eq. (\ref{ebviii}).

To investigate the critical behavior of (\ref{ebxiii}) in more detail, we generalize
(\ref{ebxiv}) to arbitrary dimension $D$.  Simultaneously,
we restore in the action the residual part $V_{\rho}$ (\ref{ebix})
of the contact interaction between loops and s vortex lines,
i.e., the part not already incorporated
into the values of coupling constants appearing in (\ref{ebxiv}):
\begin{equation}
\int d^Dr\bigl[m_{\Phi}^2|\Phi_d |^2 + |(\partial_{\mu} -iA_{\mu})\Phi _d|^2
+\frac{1}{2} g_0|\Phi_d|^4+ g_{\rho}|\Phi_d |^2J_0 +
iJ_{\mu}A^{\mu} +\frac{1}{2M^2_{\mu}}J_{\mu}J^{\mu}+
{\cal W}(J_{\mu})
+\frac{1}{2q^2_d}F_{\mu\nu}F^{\mu\nu}\bigl]~~.
\label{ebxv}
\end{equation}
Here $g_{\rho}$ denotes $V_{\rho}$ rescaled to its canonical dimension, 
$\mu =0,1,2,\dots, D-1$,
${\cal W}(J_{\mu})$ is the generalization of ${\cal W}(\delta\rho,\vec j)$,
and $F_{\mu\nu}=\partial_{\mu}A_{\nu}-\partial_{\nu}A_{\mu}$.
The functional integration runs over fields
$\Phi_d$, $A_{\mu}$ and $J_{\mu}$ and includes a constraint
$\partial_{\mu} J^{\mu}=0$. The integration over $J_{\mu}$
generates a finite mass $M_{\mu =0}=M_{\parallel}$, $M_{\mu \not=0}=M_{\perp}$ 
for the dual gauge field $A_{\mu}$.
The theory, however, retains gauge invariance 
implying that the combination $q_dA_{\mu}$ must be an RG invariant.
This in turn sets the canonical dimension of 
$g_{\rho}$ to $[g_{\rho}]=$ (length)$^{D-3}$.
Therefore, the term $g_{\rho}|\Phi_d|^2J_0$ is 
{\em irrelevant} within $\epsilon$ expansion
around the upper critical dimension $D=4$.
Similarly, all higher powers and derivatives of $J_{\mu}$ appearing
in ${\cal W}$, are irrelevant as well. For example, the canonical dimension
of the $J_0^3$ coupling constant is (length)$^{2D-3}$.
The relevant couplings below $D=4$ are $g_0$, $q_d^2$ and $M_{\parallel,\perp}$. Due to finite
$M_{\parallel,\perp}$, $A_{\mu}$ decouples and the $\beta$ function for $g_0$ is
the same as that of the neutral ($q_d=0$) complex $\Phi^4$-theory. 
In this case the $\epsilon$ expansion is expected to hold down to, and include,
$D=3$, where the critical behavior of the $\Phi^4$-theory should be that of a (inverted)
3D XY model, in agreement with our earlier assertion. 
The same conclusions can also be reached 
within the perturbative RG in fixed dimension $D=3$. Here 
$g_{\rho}$ is marginal at the ``engineering" level, 
the gauge field again decouples due to finite $M_{\parallel,\perp}$, 
and we can compute the relevant
$\beta$ functions at the one-loop order and to the leading order in $g_{\rho}$:
\begin{equation}
\beta_{0} (\hat g_0,\hat g_{\rho})
\equiv \frac{d\hat g_{0}}{d\log (p)}=
-\hat g_0 + {\cal C}_1\hat g_0^2~~~;
~~~\beta_{\rho}(\hat g_0,\hat g_{\rho})
\equiv \frac{d\hat g_{\rho}}{d\log (p)}=
{\cal C}_2\hat g_0\hat g_{\rho}~~~,
\label{ebxvi}
\end{equation}
where $\hat g_{0,\rho}(p)$ are the dimensionless running coupling constants and
${\cal C}_{1,2}$ are (regularization dependent) numerical constants which
are {\em both} positive,
${\cal C}_{1,2}>0$. At the (inverted) 3D XY critical point 
$\hat g_0 = 1/{\cal C}_1$ and therefore $\beta_{\rho}>0$,
indicating stability of our assumed $g_{\rho}=0$ fixed point against
residual $g_{\rho}$ perturbation.
The above results allow us to conclude that the critical theory 
(\ref{ebxiv}) (with ${\cal W}=0$) remains valid 
and that the effects of $V_{\rho}$ can be included by a proper choice of relevant
couplings, as originally assumed. The presence of
long-range interactions between vortices, mediated by ${\bf A}_d$, 
is essential for the validity of this argument.

One step remains: as $T\to T_{\Phi}(H)$, some overhangs attached to s vortex lines
become very large and we might doubt the accuracy of the straightforward non-relativistic
boson analogy approximations below Eq. (\ref{ebx}). However, throughout the
$\Phi$-ordered state, the s vortices remain ``massive" and there should always
exist a suitably defined quantum system of non-relativistic 2D bosons whose long 
``distance" ($x,y$) and ``imaginary time" ($z$) behavior faithfully emulates that of
s vortices. We therefore expect that the overall {\em symmetry} of
(\ref{ebxiii}) remains preserved at $T_{\Phi}(H)$. This leads to the generalization
of (\ref{ebxiii}):
$$Z(H)\to\sum_{N^{(0)}=0}^{\infty}\frac{1}{N^{(0)}!}
\prod_{l_0=1}^{N^{(0)}} \oint {\cal D}_b{\bf x}_{l_0}[s_{l_0}]
\int {\cal D}{\bf V}({\bf r})
\int {\cal D}{\bf A}_d({\bf r})
\exp (-S)~~~;~~~ S = \sum_{l_0=1}^{N^{(0)}} \frac{\tilde E_c}{T}\int ds_{l_0}+$$
\begin{equation}
+\int d^3r\bigl\{
\frac{\tilde V_0}{2T}
d_0^2({\bf r}) +
i{\bf n}_0\cdot {\bf A}_d +
i{\bf V}\cdot {\bf A}_d + 
\frac{2\pi ^2K_{\parallel}}{T}{\bf V}_{\parallel}^2 +
\frac{2\pi ^2K_{\perp}}{T}{\bf V}_{\perp}^2 +
\frac{1}{2e^2_d}(\nabla\times{\bf A}_d)^2\bigr\}~~,
\label{ebxvii}
\end{equation}
where ${\bf V}({\bf r})$ describes long distance ($\gg\ell$) fluctuations
of s vortex ``currents" ${\bf n}_s ({\bf r})$ (\ref{ebiv},\ref{ebv})
and satisfies $\nabla\cdot {\bf V}=0$. 
$K_{\parallel,\perp}/T$ now play the role of $m_sc_s^2/n_s$ and $m_s/n_s$
in (\ref{ebxiii}) and fully include the effect of overhang configurations
as $T\to T_{\Phi}(H)$.  At present, we cannot compute 
$K_{\parallel,\perp}(T,H)$ (nor $\tilde E_c$ and $\tilde V_0$) from first principles. 
This would require an analytic solution to the problem of large overhangs,
something far beyond the scope of this paper.
However, if we start with the general form (\ref{ebxvii}),
we can determine various parameters appearing there by connecting them self-consistently
to directly (numerically or experimentally) measurable physical
quantities. For example, $K_{\parallel,\perp}(T,H)$ can be extracted
from the components of the helicity
modulus tensor (Appendix B) or the fluctuation conductivity (Sec. VI).
We should therefore consider (\ref{ebxvii}) a self-consistent, perturbative
RG description of the $\Phi$-transition.

We can now enforce the constraint $\nabla\cdot {\bf V}=0$
by introducing a gauge field ${\bf S}({\bf r})$:
$2\pi {\bf V}\to \nabla\times {\bf S}$, $\nabla\cdot {\bf S}=0$. 
Alternatively, we can integrate over 
${\bf V}$, obtain the mass term for ${\bf A}_d$, and then decouple
it by introducing ${\bf S}$. The final result is:
$$Z(H)\to\sum_{N^{(0)}=0}^{\infty}\frac{1}{N^{(0)}!}
\prod_{l_0=1}^{N^{(0)}} \oint {\cal D}_b{\bf x}_{l_0}[s_{l_0}]
\int {\cal D}{\bf S}({\bf r})
\int {\cal D}{\bf A}_d({\bf r})
\exp (-S)~~~;~~~ S = \sum_{l_0=1}^{N^{(0)}} \frac{\tilde E_c}{T}\int ds_{l_0}+$$
\begin{equation}
+\int d^3r\bigl\{
\frac{\tilde V_0}{2T}
d_0^2({\bf r}) +
i{\bf n}_0\cdot {\bf A}_d +
\frac{i}{2\pi}(\nabla\times {\bf S})\cdot {\bf A}_d +
\frac{K_{\parallel}}{2T}(\nabla\times {\bf S})_{\parallel}^2 +
\frac{K_{\perp}}{2T}(\nabla\times {\bf S})_{\perp}^2
+\frac{1}{2e^2_d}(\nabla\times{\bf A}_d)^2\bigr\}~~.
\label{ebxviii}
\end{equation}
This is just the vortex loop expansion (\ref{ebvi},\ref{ebvii}) 
of the Meissner phase of a ``superconductor" 
described by an order parameter $\Phi ({\bf r})$ and coupled to the
gauge field ${\bf S}$ ($\nabla\cdot {\bf S} =0$). 
The GL functional of such a superconductor is
\begin{equation}
{\cal F}_{\rm eff} = 
\tilde\alpha\vert\Phi\vert ^2 + \tilde\gamma_{\mu}
\vert (\nabla_{\mu} + i{\bf S}_{\mu})\Phi\vert^2
+ \frac{\tilde\beta}{2}\vert\Phi\vert^4 +
\frac{K_{\parallel}}{2}(\nabla\times {\bf S})_{\parallel}^2 +
\frac{K_{\perp}}{2}(\nabla\times {\bf S})_{\perp}^2~~~,
\label{ebxix}
\end{equation}
which is precisely the gauge theory (\ref{eiii}).
$\tilde\alpha$, $\tilde\beta$ and $\tilde\gamma_{\mu}$ are some suitably
renormalized GL coefficients which can be determined phenomenologically.
Note that we have now restored the anisotropy in $\tilde\gamma_{\mu}$,
arising both from the bare anisotropy ($\gamma_{\parallel}\not =
\gamma_{\perp}$ in Eq. (\ref{ei})) and 
the anisotropy induced by the interaction
of loops with the s vortex background.


\vskip .15in
\section*{Appendix B: Gauge Theory and the Helicity Modulus}
Here we consider the connection between $K_{\perp,\parallel}$ appearing in Eq. (\ref{eiii})
and the helicity modulus tensor $\Upsilon ({\bf q})$. The conventional
definition of the components of $\Upsilon ({\bf q})$ can be found in Ref.\cite{teitel}:
\begin{equation}
\Upsilon_{\mu\nu} ({\bf q}) =V\frac{\delta^2 F}{\delta {\bf a}_{\nu}({\bf q})
\delta {\bf a}_{\mu} (- {\bf q})}~~~. 
\label{eai}
\end{equation}
All quantities appearing in (\ref{eai}) 
are defined in Sec. V, below Eq. (\ref{exvii}). We will limit our attention 
to the isotropic case ($\gamma_{\perp}=\gamma _{\parallel}$ in Eq. (\ref{ei}))
whenever we consider the $H\not =0$ situation. The
anisotropic case with finite field
requires far more extensive algebra and can be reconstructed
by combining the discussion below with illuminating presentation in Ref.\cite{teitel}.

We first evaluate $\Upsilon ({\bf q}) $ from the original GL theory (\ref{ei})
and start with the $H=0$ (${\bf A} =0$) case. We consider the situation right {\em above}
$T_{c0}$, so there is no superconducting long range order.
After adding small ${\bf a}$ to the gradient term, we can expand the free
energy up to second order in ${\bf a}$:
\begin{equation}
F[\nabla\times {\bf a}] = F[0] + \int d^3r\Bigl[
\frac{e^2}{2c^2}\chi _{\perp} (\nabla\times{\bf a})_{\perp}^2 +
\frac{e^2}{2c^2}\chi _{\parallel} (\nabla\times{\bf a})_{\parallel}^2\Bigr]
 +\dots~~~.
\label{eaii}
\end{equation}
This equation requires a brief explanation: $e$ is the real electric 
charge and $c$ is the real speed of light, appearing in (\ref{ei}).
$F[0]$ is just the original free energy of the GL theory with $H=0$ {\em and}
${\bf a}=0$. The full free energy with small ${\bf a} ({\bf r})$ is
written as a functional of $\nabla\times {\bf a}$ {\em only}. This is required by
the gauge invariance of (\ref{ei}). Two additional terms, proportional to the 
perpendicular and parallel components of $(\nabla\times {\bf a})^2$, represent
the leading corrections in powers and derivatives of $\nabla\times {\bf a}$. The
subleading contributions are denoted by dots. These subleading terms are unimportant
in the long wavelength limit. 

$\chi _{\perp}$ and $\chi _{\parallel}$ are some functions of $T$ and are different
from each other in the anisotropic case, $\gamma _{\parallel}\not =\gamma_{\perp}$,
while $\chi _{\perp} =\chi _{\parallel}=\chi$ if the superconductor is isotropic.
As $T$ approaches $T_{c0}$ from above, these functions take the following form:
\begin{equation}
\chi _{\perp} = {\cal C}T\xi _{\parallel}~~~~,~~~
\chi _{\parallel} = {\cal C}T\frac{\xi_{\perp}^2}{\xi _{\parallel}}~~~.
\label{eaiii}
\end{equation}
$\xi _{\perp,\parallel} = \xi_{0\perp,\parallel}t^{-\nu}$ are the 
superconducting correlation lengths and ${\cal C}$ is an unknown universal constant,
intrinsic to the GL theory (\ref{ei}). Note that $e^2\chi _{\perp,\parallel}/c^2$
is just the perpendicular (parallel) magnetic susceptibility.

The components of the helicity modulus tensor in the long wavelength
limit ($q\to 0$)  are uniquely determined by 
$\chi _{\perp,\parallel}$. Conversely, the measurement of the long wavelength
``tilt" and ``compression" helicity moduli\cite{teitel}
determines $\chi _{\perp,\parallel}$. In general, from definition (\ref{eai}),
we find:
\begin{equation}
\frac{c^2}{e^2}\Upsilon _{\mu\nu}({\bf q}) = 
\chi \epsilon _{\rho\alpha\mu}\epsilon _{\rho\beta\nu}q_{\alpha}q_{\beta}~~~,
\label{eaiv}
\end{equation}
for the isotropic case, while for the anisotropic situation
$\chi _{\perp}\not =\chi_{\parallel}$:
\begin{equation}
\frac{c^2}{e^2}\Upsilon _{\mu\nu}({\bf q}) = 
(\chi_{\parallel} -\chi _{\perp}) 
\epsilon _{z\alpha\mu}\epsilon _{z\beta\nu}q_{\alpha}q_{\beta}
+
\chi _{\perp} \epsilon _{\rho\alpha\mu}\epsilon _{\rho\beta\nu}q_{\alpha}q_{\beta}~~~.
\label{eav}
\end{equation}
$\epsilon _{\alpha\beta\gamma}$ is the Levi-Civita symbol and
summation over repeated indices is understood. $e^2/c^2$ is factored out 
for later convenience.

After this preliminary discussion, we go to the case of interest, finite
$H$ in Eq. (\ref{ei}),  and limit our consideration to the isotropic situation
$\gamma _{\perp}=\gamma _{\parallel}$. We introduce small ${\bf a}$ into 
Eq. (\ref{ei}) and expand to second order in $\nabla\times {\bf a}$:
\begin{equation}
F[H,\nabla\times {\bf a}] = F[H,0] + \int d^3r\Bigl[
\frac{e^2}{2c^2}\tilde\chi _{\perp} (\nabla\times{\bf a})_{\perp}^2 +
\frac{e^2}{2c^2}\tilde\chi _{\parallel} (\nabla\times{\bf a})_{\parallel}^2\Bigr]
 +\dots~~~,
\label{eavi}
\end{equation}
where $F[H,0]$ is now the free energy of the GL theory (\ref{ei}) at {\em finite}
field ${\bf H}$ and ${\bf a}=0$. We are again focusing on the ``normal", i.e. not
{\em superconducting} state, in accordance with assumption ii) of Sec. II.
The above expression looks very much like (\ref{eaii}) but there are
following significant differences: the expansion is anisotropic,
$\tilde\chi _{\perp}\not =\tilde\chi _{\parallel}$, even though 
our superconductor is isotropic. The reason for this is finite field
${\bf H}$ along the z-axis which reduces spherical symmetry of the
$H=0$ situation down to cylindrical. The finite field also breaks the ``up-down"
symmetry along the z-axis. This is manifested by the subleading corrections,
denoted by dots in (\ref{eavi}), containing in general {\em odd} powers of
$\nabla\times {\bf a}$ (the leading such term is cubic); such terms
were prohibited by symmetry in the $H=0$ case. Again, by combining
definition (\ref{eai}) and Eq. (\ref{eavi}), we arrive at the expression
for the long wavelength limit of the helicity modulus:
\begin{equation}
\frac{c^2}{e^2}\Upsilon _{\mu\nu}({\bf q}) = 
(\tilde\chi_{\parallel} -\tilde\chi _{\perp}) 
\epsilon _{z\alpha\mu}\epsilon _{z\beta\nu}q_{\alpha}q_{\beta}
+
\tilde\chi _{\perp} \epsilon _{\rho\alpha\mu}\epsilon _{\rho\beta\nu}q_{\alpha}q_{\beta}~~~.
\label{eavii}
\end{equation}

The components of the helicity modulus tensor for the finite field (isotropic)
case are determined by $\tilde\chi _{\perp,\parallel}$ which are some functions
of $T$ and $H$. It is tempting to conclude that:
\begin{equation}
\tilde\chi _{\perp} \to {\cal C}T\xi _{\parallel}(T,H)~~~~,~~~
\tilde\chi _{\parallel} \to {\cal C}T\frac{\xi^2_{\perp}(T,H)}{\xi _{\parallel}(T,H)}~~~,
\label{eaviii}
\end{equation}
where $\xi _{\perp,\parallel}(T,H)$ are now superconducting correlation lengths
at finite field. This result is plausible on physical grounds, expressing the
fact that, with $H\not =0$, the superconducting correlation lengths are now
finite in the ``liquid" phase, limited by magnetic length $\ell$ 
and consequently the helicity
moduli vanish in the $q\to $ limit. We are simply making the assumption that
the same length that limits the range of superconducting correlations appears
in the coefficient of the $q^2$-term in the helicity modulus; this assumption
is known to be correct for the $H=0$ case (\ref{eaiii}). Unfortunately, I am unable
to provide a mathematical proof that the conjectured result (\ref{eaviii}) is exact.
Instead, I {\em define} perpendicular and parallel ``screening" lengths
$\Lambda_{\perp} (T,H)$ and $\Lambda _{\parallel}(T,H)$ by:
\begin{equation}
\tilde\chi _{\perp} =   {\cal C}T\Lambda _{\parallel}(T,H)~~~~,~~~
\tilde\chi _{\parallel} =   {\cal C}T\frac{\Lambda^2_{\perp}(T,H)}
{\Lambda _{\parallel}(T,H)}~~~.
\label{eaix}
\end{equation}
Note that there is a one-to-one correspondence between these ``screening" lengths
$\Lambda _{\perp,\parallel}$
and $\chi _{\perp,\parallel}$ and, in turn, between 
$\Lambda _{\perp,\parallel}$ and the long wavelength helicity moduli (\ref{eavii}).
Since $\Lambda _{\perp,\parallel}$ are purely thermodynamic quantities they satisfy
scaling laws of Sec. IV, just like the superconducting  correlation lengths:
\begin{equation}
\Lambda _{\perp,\parallel}(T,H) = \ell {\cal R}_{\perp,\parallel}^{\Lambda}
(q^2_0)~~~~,~~~
\xi _{\perp,\parallel}(T,H) = \ell {\cal R}_{\perp,\parallel}^{\xi}
(q^2_0)~~~~,
\label{eax}
\end{equation}
where ``dimensionless charge" 
$q^2_0 = \xi (T,H=0)/\ell \propto H^{1/2}/|t|^{\nu}$ is our scaling variable of
Eq. (\ref{exv}) and 
${\cal R}_{\perp,\parallel}^{\Lambda}$ and 
${\cal R}_{\perp,\parallel}^{\xi}$ are the screening length and correlation
length scaling functions, respectively. In the $H\to 0$ ($q_0^2\to 0$) limit, all these
scaling functions, 
${\cal R}_{\perp,\parallel}^{\Lambda} (q_0^2)$ and 
${\cal R}_{\perp,\parallel}^{\xi} (q_0^2)$, go as $q_0^2$. We now make the following
{\em assumption}: around $T_{\Phi}(H)$, the ratios
\begin{equation}
\frac{\Lambda _{\perp}(T,H)}{\xi_{\perp}(T,H)} = \frac{{\cal R}_{\perp}^{\Lambda}(q^2_0)}
{{\cal R}_{\perp}^{\xi}(q_0^2)}~~~~,~~~
\frac{\Lambda _{\parallel}(T,H)}{\xi_{\parallel}(T,H)} = \frac{
{\cal R}_{\parallel}^{\Lambda}(q^2_0)}
{{\cal R}_{\parallel}^{\xi}(q_0^2)}~~~~,
\label{eaxi}
\end{equation}
are some unremarkable {\em smooth} functions of the scaling variable $q_0^2$.
In particular, the non-analytic drop in the coefficient of the $q^2$-term in
the helicity modulus, which takes place as we cross the $T_{\Phi}(H)$ transition
line (\ref{exix}) and which is directly reflected as a non-analytic
decrease in the screening lengths $\Lambda _{\perp,\parallel} (T,H)$, is 
manifested also
in the superconducting correlation lengths $\xi _{\perp,\parallel} (T,H)$.
This assumption, which appears justified physical grounds, was used in the discussion
of fluctuation conductivity (Sec. VI).

Finally, we are in position to discuss our anisotropic gauge theory of
Eq. (\ref{eiii}). Small ${\bf a}$ added to the external vector potential 
${\bf A}$ in the original GL theory (\ref{ei}) translates into
a small vector potential $(e/c){\bf a}$ added to our fictitious gauge
field ${\bf S}$ in Eq. (\ref{eiii}). Since we are integrating over ${\bf S}$,
it is useful to define new gauge field ${\bf S_n}={\bf S} + (e/c){\bf a}$
and integrate over ${\bf S_n}$ in the partition function. The effect
of this is to move $(e/c){\bf a}$ from the covariant gradient terms,  $|D_{\mu}\Phi |^2$,
to the ``gauge field energy" $K_{\mu}[\nabla\times ({\bf S}- (e/c){\bf a})]^2_{\mu}$
in Eq. (\ref{eiii}). We now
expand the free energy of the gauge theory, $F_{\rm eff}$, to second order
in $\nabla\times {\bf a}$, following the same philosophy as in Eq. (\ref{eavi}):
\begin{equation}
F_{\rm eff}[\tilde e_{\perp},\tilde e_{\parallel},\nabla\times {\bf a}] = 
F_{\rm eff}[\tilde e_{\perp},\tilde e_{\parallel},0] + \int d^3r\Bigl[
\frac{e^2}{2c^2}{\cal K}_{\perp} (\nabla\times{\bf a})_{\perp}^2 +
\frac{e^2}{2c^2} {\cal K}_{\parallel} (\nabla\times{\bf a})_{\parallel}^2\Bigr]
 +\dots~~~,
\label{eaxii}
\end{equation}
where
\begin{equation}
{\cal K}_{\perp,\parallel} = K_{\perp,\parallel} - \frac{K_{\perp,\parallel}^2}{T}
\lim_{q\to 0^{+}}
\int d^3 ({\bf r}-{\bf r'})e^{i{\bf q}\cdot ({\bf r}-{\bf r'})}
\langle (\nabla\times {\bf S}({\bf r}))_{\perp,\parallel}
(\nabla'\times {\bf S}({\bf r'}))_{\perp,\parallel}\rangle~~~.
\label{eaxiii}
\end{equation}
The anisotropic charges $\tilde e_{\perp,\parallel} (T,H)$ and coupling constants
$K_{\perp,\parallel}(T,H)$ are defined in Eqs. (\ref{exiii})
and (\ref{eiii}), respectively.  
The thermal average $\langle\cdots\rangle$ is over the gauge theory defined
by the free energy functional ${\cal F}_{\rm eff}$ (\ref{eiii}).
Combining (\ref{eai}) and (\ref{eaxiii}), we get, as before:
\begin{equation}
\frac{c^2}{e^2}\Upsilon _{\mu\nu}({\bf q}) = 
({\cal K}_{\parallel} -{\cal K}_{\perp}) 
\epsilon _{z\alpha\mu}\epsilon _{z\beta\nu}q_{\alpha}q_{\beta} +
{\cal K}_{\perp} \epsilon _{\rho\alpha\mu}\epsilon _{\rho\beta\nu}q_{\alpha}q_{\beta}~~~.
\label{eaxiv}
\end{equation}
Comparing this to the general expression for the helicity modulus of the GL theory
at finite field, given by Eq. (\ref{eavii}), we conclude that
$\tilde\chi _{\perp,\parallel}={\cal K}_{\perp,\parallel}$.

A {\em very} important point:
{\em below} $T_{\Phi}(H)$ we have ${\cal K}_{\perp,\parallel}=
K_{\perp,\parallel}$. 
This is a mathematical consequence of the following physical 
picture (Sec. V). In the $\Phi$-ordered state, {\em only} the field-induced (s) vortex lines
can ``screen" the test vector potential ${\bf a}({\bf r})$. All thermally generated
vortex loops are of {\em finite} size and cannot contribute to the ``screening"
in the long wavelength limit, described by ${\cal K}_{\perp,\parallel}$
in Eq. (\ref{eaxii}). This implies that the new order parameter $\Phi$ is
{\em finite} and therefore 
$\langle (\nabla\times {\bf S})^2_{\perp,\parallel}\rangle$ (\ref{eaxiii})
{\em vanishes} in the long wavelength limit, as 
$\sim q^2/|\langle\Phi\rangle|^2$. Therefore,
we can {\em uniquely fix} the coupling constants 
$K_{\perp,\parallel}(T,H)$ 
(and corresponding anisotropic charges $\tilde e_{\perp,\parallel} (T,H)$)
that enter the gauge theory description (\ref{eiii}), by connecting them directly
to the components of the helicity modulus tensor
right below $T_{\Phi}(H)$ (or, more precisely, for
$T_{\Phi}(H) - T\to 0^+$).  In particular, the functions $c_{\perp,\parallel}(T,H)$,
introduced below Eq. (\ref{eiii}) and referred to at various points in
the main text, follow from the general expression (\ref{eaix}):
\begin{equation}
c_{\perp}(T,H)= {\cal C}{\cal R}_{\parallel}^{\Lambda}(q_0^2)~~~,
~~~c_{\parallel}(T,H)= {\cal C}\frac{({\cal R}_{\perp}^{\Lambda}(q_0^2))^2}
{{\cal R}_{\parallel}^{\Lambda}(q_0^2)}~~~,
\label{eaxv}
\end{equation}
where ${\cal R}_{\perp,\parallel}^{\Lambda} (q_0^2)$ are the 
scaling functions introduced in Eq. (\ref{eax}) and are to be evaluated 
below $T_{\Phi}(H)$. The gauge theory scenario predicts that the
{\em fundamental anisotropy}
ratio $c_{\parallel}/c_{\perp}$ takes on a universal value along $T_{\Phi}(H)$.

Above $T_{\Phi}(H)$, in the true normal state, infinite vortex loops proliferate across
the system and {\em can} contribute to screening.     
$\langle (\nabla\times {\bf S})^2_{\perp,\parallel}\rangle$ in Eq. (\ref{eaxiii})
becomes finite and $\sim 1/\xi _{\Phi}$. This causes 
the non-analytic drop in the $q^2$-term 
of the helicity modulus and the corresponding screening
lengths (\ref{eaix}), just as discussed in Sec. V (see Eq. (\ref{exix})).

The reader should note that the set of results presented in
this Appendix, connecting the properties of the gauge
theory description (\ref{eiii}) around $T_{\Phi}(H)$
to the general long distance form of the helicity modulus and 
screening lengths of the GL theory at finite field (\ref{ei}), is not only 
physically transparent and appealing but also exact,
provided our assumptions i)-iv) (Sec. II), are satisfied.

\newpage
\begin{figure}
\epsfxsize=7.1cm
\vskip 0.3in
\caption[]{Proposed $H-T$ phase diagram for the critical region
of extreme type-II superconductors. The dashed region denotes a
crossover from the Gaussian regime, where amplitude fluctuations
are strong, to the critical 3D XY-like regime, where amplitude
fluctuations are suppressed. Within the 3D XY-like critical regime,
the London-type vortex loops and lines with tight cores are well-defined excitations. 
Along the temperature axis, this critical
region is bounded by the {\em mean-field} $T_c$. Along the field axis,
the critical region is bounded by $H_s$ ($\sim H_b$) (\ref{ev}). 
Above $H_s$, the physics of GL theory (\ref{ei}) is dominated by the
formation of Landau levels for Cooper pairs.\cite{zta,hu}
The $\Phi$-transition, or the vortex-loop ``expansion" transition, $T_{\Phi}(H)$,
and the vortex lattice melting, $T_m(H)$, occur {\em simultaneously}
for $H>H_Z\sim H_s$. This is a {\em first-order} transition from the
Abrikosov vortex lattice (VS) directly to the normal state (N).  
Below $H_Z$, $T_{\Phi}(H)$ and $T_m(H)$ split into two {\em separate} transitions
and merge again only at the true zero-field
superconducting transition, $T_{c0}$, as $H\to 0$. For $H\ll H_s$, the transition
at $T_{\Phi}(H)$ is {\em continuous}, while the vortex lattice melting transition
remains first-order. The intermediate phase ($\Phi$), below $T_{\Phi}(H)$
but above $T_m(H)$, is not a superconductor ($\langle\Psi\rangle =0$), but it differs
from the true normal state (N) by a new type of long range order, characterized
by the ``superconducting" order parameter $\Phi ({\bf r})$ (\ref{eii},\ref{eiii}).
Only $N_{\Phi}$ field-induced vortex lines traverse the system along the field
direction in this $\Phi$-ordered state, while the average size of thermally-generated
vortex loops is finite. In the true normal state (N), the $\Phi$-order is
destroyed as numerous {\em additional} vortex paths ``expand"
across the system in all directions.
}
\label{fig1}
\end{figure}

\begin{figure}[t]
\epsfxsize=7.1cm
\vskip 0.3in
\caption[]{A schematic representation of the $H=0$ transition in an
extreme type-II superconductor. The low temperature Meissner phase 
($T<T_{c0}$) contains only finite vortex loops. In the high temperature
normal state ($T>T_{c0}$) these loops connect and ``expand"
across the system, leading to loss of phase coherence and finite
dissipation. Two forms of superconducting order, described by $\Psi$ and $\Phi$,
are equivalent here. For clarity, the vortex paths are drawn far smoother
than they actually are near $T_{c0}$.
}
\label{fig2}
\end{figure}

\begin{figure}[t]
\epsfxsize=7.1cm
\vskip 0.3in
\caption[]{Characteristic configurations of the system a) below and
b) above $T_{\Phi}(H)$ (but always above $T_m(H)$). a) Field-induced
vortices (depicted in blue) wind all the 
way across the system along the field direction
but only undergo effective ``diffusion" in the transverse direction.
b) After the loop ``expansion" at $T_{\Phi}(H)$ this effective
transverse ``diffusion" is destroyed, as field-induced vortices can
``hitch" a ride all the way across the system in the xy-plane 
by connecting to thermally-generated infinite 
loops present in the true normal state.
Note the presence of ``vortex tachyons" (depicted in red) which
wind only in the xy-plane. Again, for illustrative purposes,
the vortex paths are drawn far smoother
than they actually are near $T_{\Phi}(H)$.
}
\label{fig3}
\end{figure}

\begin{figure}[t]
\epsfxsize=7.1cm
\vskip 0.3in
\caption[]{Characteristic configurations of the system a) below and
b) above $T_m (H)$ (but always below $T_{\Phi}(H)$). 
a) Field-induced vortices (blue)
execute small oscillations around their equilibrium positions. Thermally-generated
vortex-loops (red) are small and rare.
b) Above the melting transition thermally-generated loops discontinuously 
grow larger and more numerous, although they still remain of finite size.
This discontinuous change in the state of 
the loops accounts for $\langle\Phi\rangle_S
\not =\langle\Phi\rangle_L$ at the simplest mean-field level.
}
\label{fig4}
\end{figure}


\begin{references}
\bibitem{helium} The main results of the present theory should also be applicable,
with appropriate modifications, to uniformly rotating superfluid helium.
\bibitem{footiii} ``Perturbed" comes with quotation marks since, even though
the magnetic field enters Eq. (\ref{eiii}) through a small parameter ${\tilde e}$,
this perturbation is strongly relevant and it immediately destabilizes the
zero-field (``neutral") 3D XY critical point.
\bibitem{reversal} Note that $T_{\Phi}(H)<T_m(H)$ is not excluded
at higher fields.
\bibitem{footv} Unlike the mean-field based approach including only the field-induced
London vortex lines, where the strength of this interaction is set by the
mean-field {\em amplitude} squared of $\Psi ({\bf r})$.
\bibitem{tphi} Eq. (\ref{eix}) for $t_{\Phi}(H)$ differs
from the estimate reported in Ref.\cite{zt} in two ways: first, the previous
estimate was made for the weakly-coupled layered case and, second, it
was based on the mean-field approximation to (\ref{eii}). In contrast,
the present expression (\ref{eix}) is exact, provided $d_{\Phi}$ is
specified.
\bibitem{charge} This fundamental anisotropy is present even for 
$\xi_{\perp} =\xi_{\parallel}$ ($\Gamma =1$) and is set by $c_{\perp}/c_{\parallel}$.
The value of $c_{\perp}/c_{\parallel}$ along the $T_{\Phi}(H)$ line 
(more precisely, for $T_{\Phi}(H) - T\to 0^+$) is a universal ratio of 
the gauge theory scenario.
\bibitem{meltingline} One naively expects the range of fields over which
$d_m$ is universal to be smaller than the corresponding range for $d_{\Phi}$:
the melting line is first-order and involves a non-uniform low temperature
state.
\bibitem{feigelman} In principle, however,
such phase could result in a new transition line between $T_{\Phi}(H)$
and $T_m(H)$, at least in a certain window of material parameters.
\bibitem{footx} The same would be true for the LL-based theories, if uncritically
extended to the $H\to 0$ limit.
\end{references}
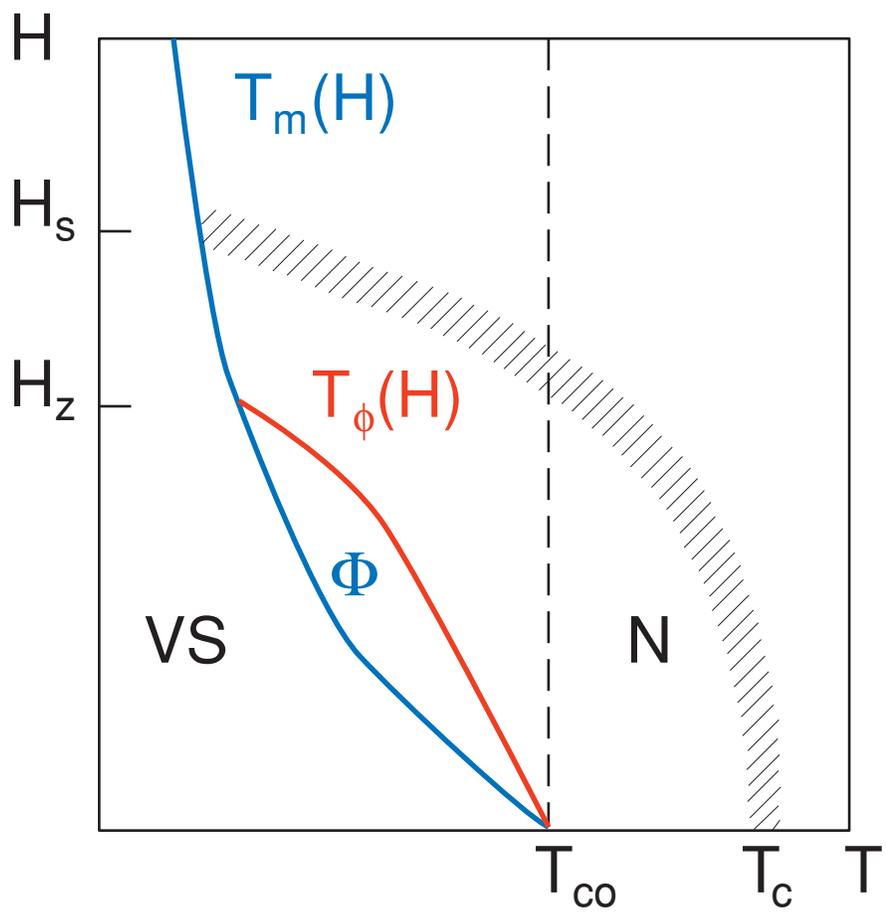
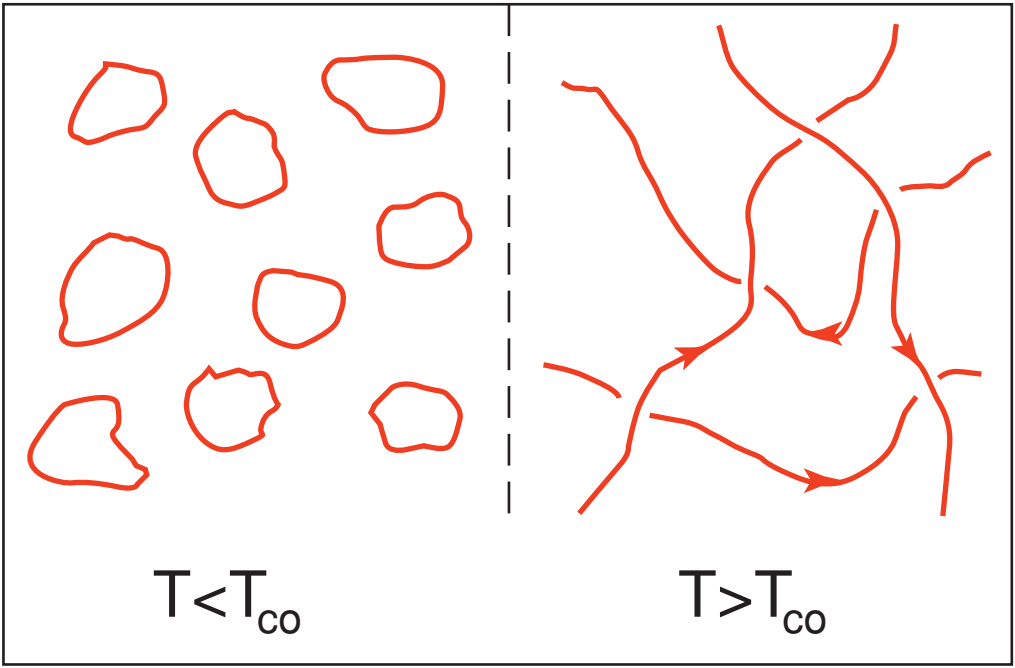
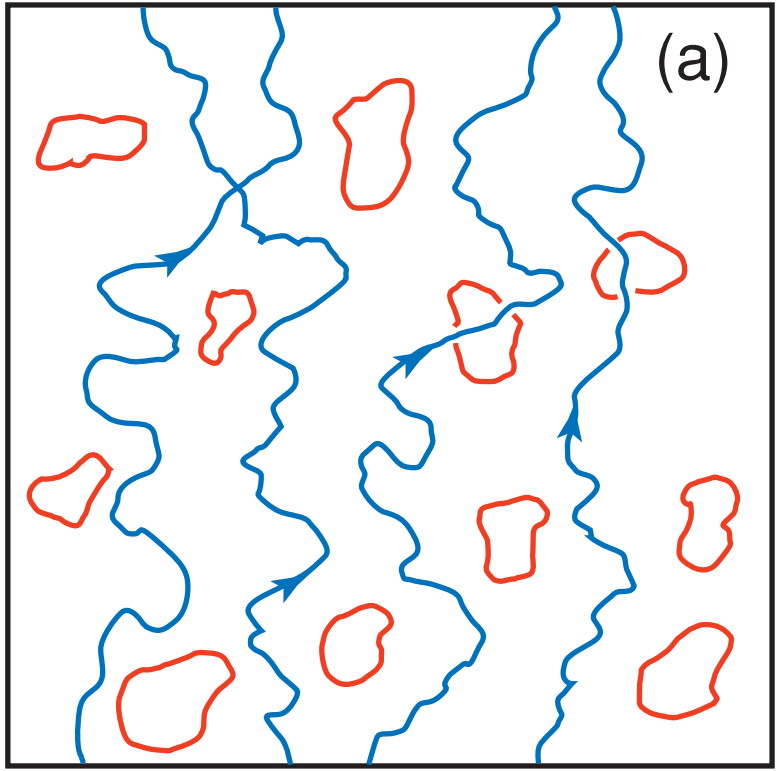
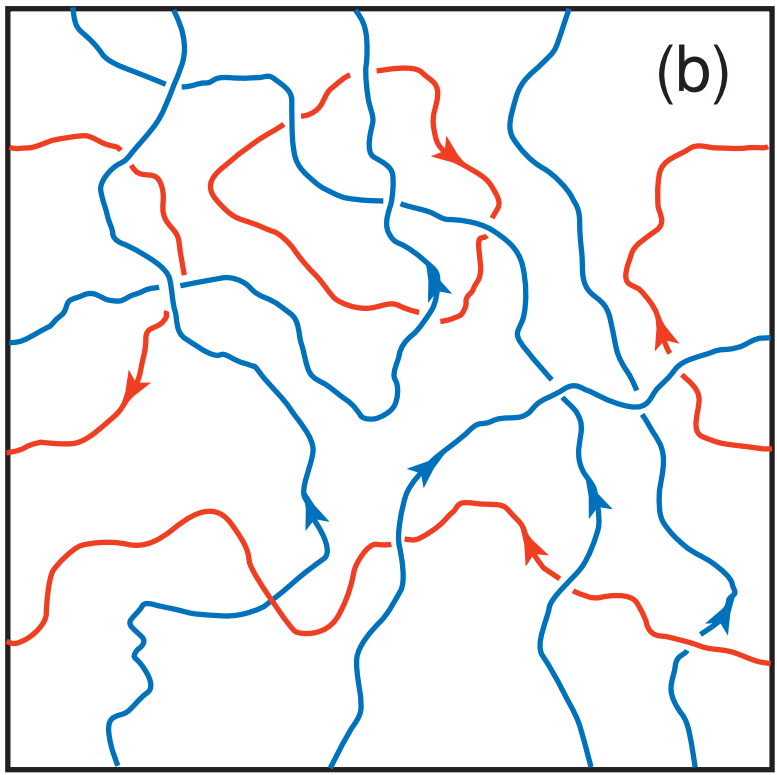
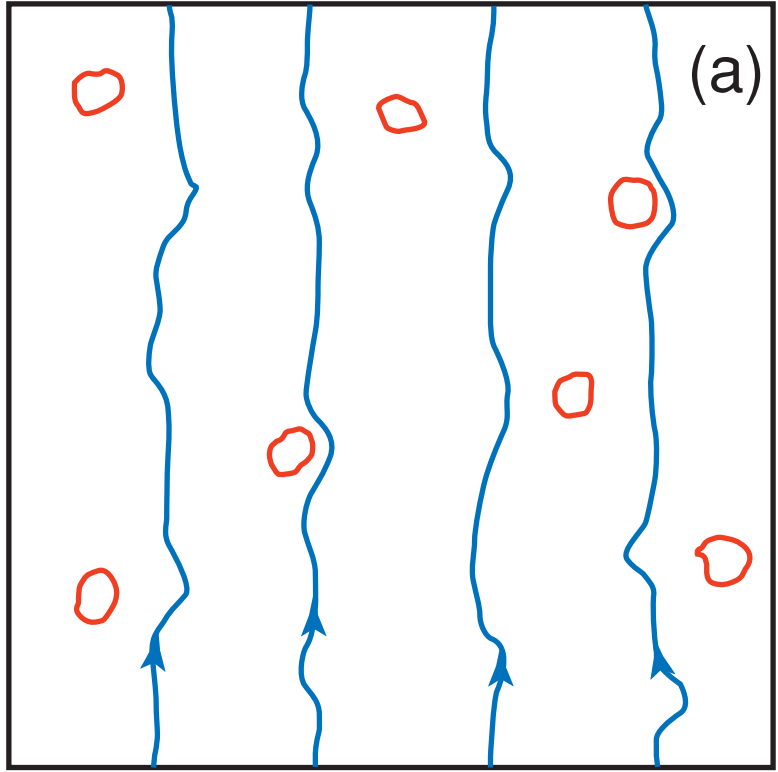
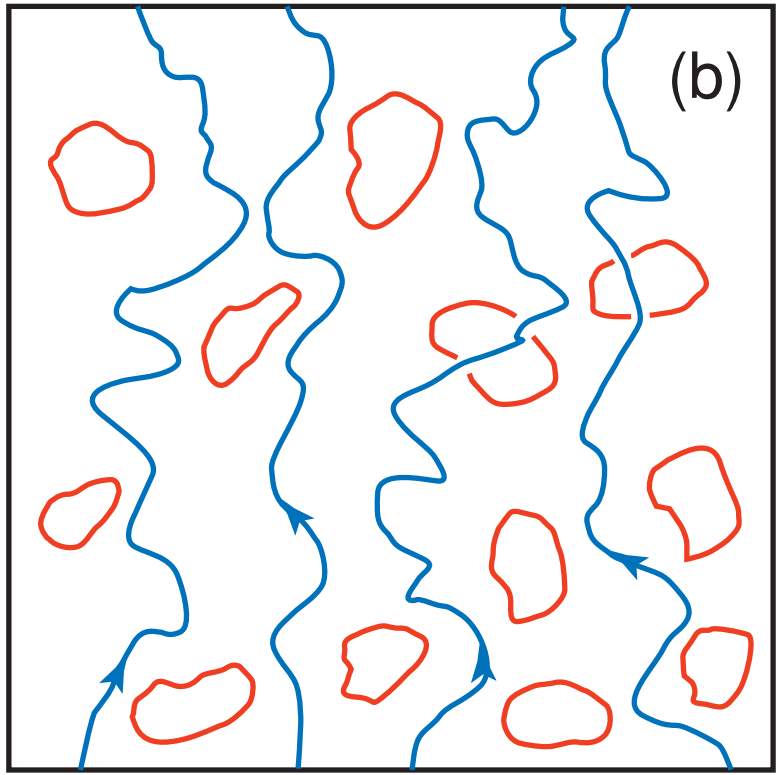
\end{document}